\documentclass[submission,copyright,creativecommons]{eptcs}
\providecommand{\event}{OCaml 2017 Post-proceedings}
\usepackage{url}

\usepackage{underscore}
\usepackage{longtable}
\usepackage{parskip}

\usepackage{graphicx}
\usepackage{pdflscape}
\usepackage{xcolor}
\usepackage[T1]{fontenc}
\usepackage{textcomp}
\usepackage[scaled=0.9]{beramono}
\usepackage{upquote}

\usepackage{fancyvrb}
\fvset{fontsize=\footnotesize}
\RecustomVerbatimEnvironment{verbatim}{Verbatim}{}
\usepackage{microtype}

\title{Direct Interpretation of Functional Programs for Debugging}
\author{
John Whitington \qquad\qquad Tom Ridge
\institute{University of Leicester\\
UK}
\email{\quad \{jw642,tr61\}@le.ac.uk}
}
\def\titlerunning{Direct Interpretation of Functional Programs for Debugging}
\def\authorrunning{John Whitington \& Tom Ridge}
\begin{document}
\maketitle

\begin{abstract}
We make another assault on the longstanding problem of debugging. After exploring why debuggers are not used as widely as one might expect, especially in functional programming environments, we define the characteristics of a debugger which make it usable and thus widely used. We present initial work towards a new debugger for OCaml which operates by direct interpretation of the program source, allowing the printing out of individual steps of the program's evaluation. We present \textsf{OCamli}, a standalone interpreter, and propose a mechanism by which the interpreter could be integrated into compiled executables, allowing part of a program to be interpreted in the same fashion as \textsf{OCamli} whilst the rest of the program runs natively. We show how such a mechanism might create  a source-level debugging system that has the characteristics of a usable debugger (such as being independent of its environment) and so may eventually be expected to be suitable for widespread adoption.
\end{abstract}

\section{Introduction}

This paper describes early work on a new approach to debugging programs written in the functional programming language OCaml \cite{ocaml}. This approach is informed by a look at the history and characteristics of debugging, a mild astonishment at the lack of use of debuggers in the functional programming community in particular, and the identification of characteristics a usable debugger must have. In addition to presenting the work which has been done, we give informed speculation on the final shape of our debugger, since the implementation is not yet complete. In this section, we look at the problem of debugging from a historical perspective, then go on to discuss debuggers for modern functional programming languages. In Section 2, we decide how to proceed -- what our principles will be -- and give  examples of how the interpreter might be used. Section 3 presents \textsf{OCamli}, an interpreter for OCaml programs, and describes its architecture. We propose a possible interface for the debugger, based upon OCaml syntax extensions. Finally, in Sections 4 and 5, we assess what has been done and discuss future work.

\subsection{The debugging problem}

As early as in 1965, surprise was expressed that the shift from machine code to assembly language to compiled languages to block-structured compiled languages, and the simultaneous vast improvements in computing power and cost, had not led to as great a reduction in the frequency or severity of bugs. Halpern, in ``Computer programming: the debugging epoch opens'' \cite{halpern1965computer}, writes \textit{``That tendency to err that programmers have been noticed to share with other human beings has often been treated as if it were an awkwardness attendant upon programming's adolescence, which like acne would disappear with the craft's coming of age. It has proved otherwise \ldots\ Many of us expected compiler languages to eliminate all bugs except those so glaring as to leap to the first fresh eye cast on the program. \ldots\ An unfriendly behaviorist studying programmers might conclude that we deliberately elaborate our tasks so as to keep the bug rate constant.''} \cite{halpern1965computer}. Balzer, in his 1969 description of the Rand Corporation's EXDAMS debugger \cite{balzer1969exdams}, explains that this surprise was widespread in the industry, at a time when more and more large programs were being written in modern compiled languages: \textit{``With the advent of the higher-level algebraic languages, the computer industry expected to be relieved of the detailed programming required at the assembly-language level. This expectation has largely been realised. Many systems are now being built in higher-level languages (most notably MULTICS). \ldots\ However, the ability to debug programs has advanced but little with the increased use of these high-level languages.''}
\cite{balzer1969exdams}. Hamlet, writing in 1983, gives a possible reason -- the lack of attendant progress in debugging paradigms, suggesting that high level languages might require different kinds of debugging tools rather than mere analogs of existing ones: \textit{``Debugging techniques originated with low-level programming languages, where the memory dump and interactive word-by-word examination of memory were the primary tools. `High-level' debugging is often no more than low-level techniques adapted to high-level languages.''}
\cite{hamlet1983}. It is fascinating to see Halpern, quoted earlier, writing on the same topic forty years later, in 2005: \textit{``The most remarkable thing about debugging today is how little it differs from debugging at the dawn of modern computing age, half a century ago. \ldots\ We've made little progress in debugging methods in half a century, with the result that projects everywhere are bogged down because of buggy software.''}
\cite{halpern2005assertive}

\noindent Debuggers are still not as widely used as one might expect, even in difficult domains. In a paper on debugging practices for complex legacy systems, Regelson and Anderson write: \textit{``The major item noted by survey respondents was that few people really have learned to use the capabilities of their debuggers''} \cite{regelson1994debugging}. Debugging in industry is sporadic. Parnin and Orso \cite{Parnin:2011:ADT:2001420.2001445}, writing specifically about automated debugging techniques, say: \textit{``Although potentially useful, most of these \textup{[debugging]} techniques have yet to demonstrate their practical effectiveness. One common limitation of existing approaches, for instance, is their reliance on a set of strong assumptions on how developers behave when debugging.''}
\cite{Parnin:2011:ADT:2001420.2001445}. This limitation of a ``reliance on a strong set of assumptions'' as the key to understanding why people do not use debuggers is a recurring theme. For example, Brady, in a paper about a debugging tool for experienced users \cite{brady1968writing}, explains that prior systems tried too hard to be approachable for novices, and the verbosity of their commands alienated the experienced. He writes: \textit{``In a debugging program it is of prime importance that the program be simple, flexible, and highly efficient to use.''}
 \cite{brady1968writing}. Evans and Darley, in their 1966 survey of online debugging systems, agree, explaining that when designing the interface to a breakpoint-based debugger: \textit{``Here, as in other aspects of on-line work, convenience is critical.''}
 \cite{evans1966line}. Eisenstadt \cite{eisenstadt1997my} lists three principles for the usability of a debugger:

\vspace{-6mm}
\begin{quotation}
{\noindent\itshape\begin{itemize}
\item Allow full functionality at all times. Debugging environments that prevent access to certain facilities make matters worse.
\item Viewers should be provided for ``key players'' (any evaluable expression) rather than just for ``variables''.
\item Provide a variety of navigation tools at different levels of granularity. \textup{\cite{eisenstadt1997my}}\end{itemize}}\end{quotation}

\noindent Grishman \cite{grishman1970debugging} echoes the first principle, arguing that a debugger is at its best when it is at its most widely applicable: \textit{``\ldots to facilitate maintenance, the same program was to be usable in both batch and interactive modes. Second, to facilitate distribution, the system had to be usable without any modification to the operating system.''}
\cite{grishman1970debugging}.

Let us now look at the debugging of programs written in functional languages.

\subsection{Debugging functional programs}

We should like to build a new debugger for the functional language OCaml, trying to learn lessons from the history of debugging. Some of these lessons will no doubt be language-agnostic, but we expect functional languages to have special requirements. And so, we should take a tour of existing debuggers for functional languages and examine to what extent they are usable and used.

Debuggers for functional languages have often followed the pattern of those for imperative languages, even though the mental model of evaluation as expression-rewriting is different. Concepts such as breakpoints appear often, for example. Such debuggers come in several flavours. Some work by extending low-level executable debuggers such as GDB or LLDB with extra routines to allow reconstruction of expressions, some modify the program as it is being compiled, inserting information which can be used by a specialised debugging program, and some work simply by providing macros or extra routines for debugging or logging.

Of course, most functional programming languages have a Read-Eval-Print Loop, which is used not only for learning and programmming-by-experimentation but also for light testing and debugging. Debuggers for functional languages aim to provide facilities over and above the REPL. The limitation of the REPL-as-debugger approach is that debugging often occurs due to an unexpected failure in production, rather than something the programmer provokes deliberately (which we would probably call testing). It is worth noting a practical point: in many languages, it is possible to build a REPL automatically with all libraries and modules used in a project linked in, for example by typing {\small\texttt{make\negthinspace\ repl}} instead of just {\small\texttt{make}} -- a boon for usability.

\textbf{Standard ML}\hspace{3.5mm} Refreshingly, debugging was considered, at least in passing, during the early stages of the design of Standard ML, as Hall and O'Donnell \cite{hall} quote Milner \cite{milner} recalling: \textit{``ML does \textup{not} use lazy evaluation; it calls by value. This was decided for no other reason than inability to see the consequences of lazy evaluation for debugging (remember that we wanted a language which we could use rather than research into), and the interaction with the assignment statement, which we kept in the language for reasons already mentioned.''} \cite{hall}\cite{milner}. However, the history of debugging tools for Standard ML has not always followed this pattern. Wadler \cite{Wadler} records the story of Tolmach and Appel's debugger \cite{tolmach}, which was deeply intertwined with the compiler and runtime of the SML/NJ Standard ML compiler. As the SML/NJ implementation evolved, the debugger fell out of step, and is no longer available. Standard ML developers \textit{``must return to older, more manual debugging methods''} \cite{Wadler}. This is a reminder that keeping a tool which is not part of the standard language toolchain up-to-date requires either frequent modification, or a design which is fundamentally distanced from the language. Of course the spectrum of effort required to update a debugger for a new version of the language is broad. It is likely any tool other than a GDB-style one (operating solely on executables) will always require some updating with each new toolchain release. One practical way to ensure this, if only socially, is to make it part of the toolchain.

The Poly/ML implementation of Standard ML contains an interactive debugger which operates not in a separate environment, but within the usual REPL. For example, here we set a breakpoint on a list reversal function, and ask for the values associated with some names: 

\smallskip
\begin{verbatim}[commandchars=\\\{\}]
Poly/ML 5.7.1 Release                   
> PolyML.Compiler.debug := true;         \textrm{\textit{initialise the debugger}}
val it = (): unit
> fun rev [] = []
#   | rev (h::t) = rev t @ [h];
val rev = fn: 'a list -> 'a list
> PolyML.Debug.breakIn "rev";            \textrm{\textit{enable the debugger on our function}}
val it = (): unit
> rev [1, 2, 3, 4];                      
function:rev
debug > h;                               \textrm{\textit{ask for values at the debugger prompt}}
val it = ?: 'a
debug > t;
val it = [?, ?, ?]: 'a list
\end{verbatim}
\smallskip

\noindent Notice, though, that even our simple polymorphic list reversal prevents the Poly/ML debugger from printing out the full details of the values we would like to see (one can give the type manually, but giving the wrong type can crash Poly/ML, the documentation advises). 

\textbf{F\#}\hspace{3.5mm} Microsoft's F\# \cite{fsharp} is an example of a functional language tightly integrated into (and shipped by default with) a platform of frameworks, libraries and so on, based on the Common Language Runtime \cite{clr}. Thus, we would expect F\# to be an interesting case when examining debugging functional programs in a broadly imperative scenario. The official guidance on debugging F\# \cite{fsharpdebugging} is, however, a little disheartening:

\begin{quotation}

{\itshape \noindent Debugging F\# is similar to debugging any managed language, with a few exceptions:}

\begin{itemize} 
\item \textit{The Autos window does not display F\# variables.} 

\item \textit{Edit and Continue is not supported for F\#. Editing F\# code during a debugging session is possible but should be avoided. Because code changes are not applied during the debugging session, editing F\# code during debugging will cause a mismatch between the source code and the code being debugged.}

\item \textit{The debugger does not recognise F\# expressions. To enter an expression in a debugger window or a dialog box during F\# debugging, you must translate the expression into C\# syntax. When you translate an F\# expression into C\#, make sure to remember that C\# uses == as the comparison operator for equality and that F\# uses a single =.} \cite{fsharpdebugging}
\end{itemize}
\end{quotation}

\noindent So the advantage of having a full IDE and a widely-used platform for the functional language to sit within is tempered by compromised support for debugging, at least in this case.

\textbf{OCaml}\hspace{3.5mm} We shall now look at typical methods used for debugging in the OCaml community, in addition to the use of the REPL for debugging-like tasks which we have already highlighted.

Inserting {\small\texttt{printf}} statements is a popular method of informal debugging and logging in many languages on many platforms. However, OCaml (unlike, for example, Haskell or Java), has no generic mechanism for printing user-defined data types. So one is limited to printing only parts of the data -- such as strings or numbers, or forced to use custom printers, or limited to a library whose purpose is to provide custom printers. Such restrictions can be painful. Nonetheless, inserting {\small\texttt{printf}} statements is an example of a debugging mechanism which, whilst it may not always be effective, is at least available in almost all circumstances. Perhaps it is this aspect of its usability which explains its enduring popularity.

The OCaml REPL has a very basic tracing mechanism. For example, here we define a simple function and the tracer displays inputs to and outputs from the function as it runs:

\smallskip
\begin{verbatim}[commandchars=\\\{\}]
# let rec f x = function 0 -> x | n -> f (succ x) (pred n);;
val f : int -> int -> int = <fun>
# #trace f;;                         \textrm{\textit{enable tracing for our function}}
f is now traced.
# f 0 1;;                            \textrm{\textit{invoke the function}}
f <-- 0
f --> <fun>
f* <-- 1
f <-- 1
f --> <fun>
f* <-- 0
f* --> 1
f* --> 1
- : int = 1
\end{verbatim}
\smallskip

\noindent Too much has been lost in the compilation process to provide more information about the evaluation process of the expressions. In particular, currying is not preserved. Values having polymorphic types cannot be printed but appear as {\small\texttt{<poly>}}.

The OCamldebug program is supplied with OCaml. It operates only on compiled and linked bytecode executables, not on native code executables or on source. The program must have been compiled with debug information. In addition, the build process must make a bytecode executable by default, or in addition to a native code one. The stand-out feature of OCamldebug is its ability to ``time-travel'' -- that is to jump backwards in a program's execution as well as forwards. This is achieved by the use of the unix fork mechanism. The intention is to make it easier to ``catch the bug in the act''.

The program is run in a sequence of numbered steps. A step is something like a function application or a conditional branch. One may: jump to a numbered step, forward or backward; print out the source code at the current step; inspect a value from the source code; and set breakpoints based on source code positions. As we shall see, there are some limitations. Let us take an example run. We start the debugger with the program {\small\texttt{cpdf\negthinspace\ -version}}:

\smallskip
\begin{verbatim}[commandchars=\\\{\}]
$ ocamldebug cpdf -version
	OCaml Debugger

(ocd) run
Loading program... done.
cpdf Version 2.2
Time: 49260                                         
Program exit.                        
\end{verbatim}
\smallskip

\noindent We go to time zero, the beginning of the program. We have ``time-travelled''. 

\smallskip
\begin{verbatim}[commandchars=\\\{\}]
(ocd) go 0
Time: 0
Beginning of program.
\end{verbatim}
\smallskip

\noindent We step forward one step at a time. What we see is just module initialisation from OCaml's built-in Standard Library {\small\texttt{Pervasives}}.

\smallskip
\begin{verbatim}[commandchars=\\\{\}]
(ocd) step
Time: 1 - pc: 7384 - module Pervasives
26     (Invalid_argument "index out of bounds")<|a|>
(ocd) step
Time: 2 - pc: 7552 - module Pervasives
164   float_of_bits 0x7F_F0_00_00_00_00_00_00L<|a|>
\end{verbatim}
\smallskip

\noindent We move into code from the actual program (rather than module initialization) but we are still stuck in Standard Library code, there being no way to ask OCamldebug to show only steps involving the user's main program only.

\smallskip
\begin{verbatim}[commandchars=\\\{\}]
(ocd) go 20000
Time: 20000 - pc: 136812 - module Arg
277     else <|b|>if s.[n] = ' ' then loop (n+1)
(ocd) step
Time: 20001 - pc: 136828 - module Arg
277     else if s.[n]<|a|> = ' ' then loop (n+1)
(ocd) step
Time: 20002 - pc: 136864 - module Arg
278     else <|b|>n
\end{verbatim}
\smallskip

\noindent We print some values by giving their names: 

\smallskip
\begin{verbatim}[commandchars=\\\{\}]
(ocd) print n                                      \textrm{\textit{ask for value of {\texttt{n}}}}
n: int = 1
(ocd) go 20001
Time: 20001 - pc: 136828 - module Arg
277     else if s.[n]<|a|> = ' ' then loop (n+1)
(ocd) print s                                      \textrm{\textit{ask for value of {\texttt{s}}}}
s: string = " Display this list of options"
(ocd) print loop                                   \textrm{\textit{ask for value of {\texttt{loop}}}}
Unbound identifier loop                           
\end{verbatim}
\smallskip

\noindent Some values cannot be found, or are opaque. We cannot alter the values within the debugging environment and re-run the code.

OCamldebug can be used in conjunction with the Emacs text editor \cite{stallman1981emacs} to provide for a smoother debugging experience via shortcuts for debugger commands, and the ability to jump to the source code position of a breakpoint. It is also possible to install printers for user-defined data types, although the manual cautions \textit{``For technical reasons, the debugger cannot call printing functions that reside in the program being debugged.''}

It is possible to use a debugger which works on executables (such as GDB) with OCaml, of course, but facilities are limited. The semantic gap between the source text and the executable in the functional model of computation is much wider than when debugging a language such as C. There is, however, an extension \cite{libmonda} to GDB in development, which allows for limited printing out of OCaml values using the type annotation files left behind during compilation.

\textbf{Haskell}\hspace{3.5mm} In 2005, in a survey of the users of GHC, the most prominent Haskell compiler, \textit{``By far the most common request was for a debugger''} \cite{marlow}. We shall describe three such debuggers briefly, and then discuss the results of a similar survey undertaken ten years later, in 2015. The paper describing the current debugger (shipped with GHC), says \textit{``The most prominent working debuggers for Haskell are Hat and Hood.''} \cite{marlow}, so we choose those to look at before examining the GHC debugger.

The Hat \cite{chitil2002transforming} debugger operates by recompiling programs in such a way that they dump a trace of the whole execution to file as the program runs, whether it ends normally or with an error. After the program has finished, the user runs tools which use the dumped data to explore the execution of the program. A transformed program runs about a hundred times more slowly than the original. However, Hat allows some modules (say the standard library) to be ``trusted'' and therefore untraced. This also enables Hat users to debug programs which use third-party libraries which Hat has not, or cannot, recompile.

The tools provided include {\small\texttt{hat-observe}} to show the arguments with which each function is called, {\small\texttt{hat-trail}} to explore computations backwards (to answer the question ``Where did my bug come from?''), and {\small\texttt{hat-explore}} to step through computations.

However, there are problems. The trace can be enormous, even for modest program runs. This, together with the tracing slowdown, may restrict the debugging of programs which do not fail (or otherwise end) quickly. Since Hat relies on transforming Haskell programs into ones which are semantically equivalent but which also output trace data, it cannot be used with programs which make use of language extensions Hat does not know about. Thus, one usage of a recent Haskell extension in a codebase can rule out Hat as a debugger. To debug inside libraries, Hat also requires one to recompile all the libraries in tracing variants, for use with Hat.

The Hood debugger \cite{Gill00debugginghaskell} works by printing out data structures at various ``observation points'' in the program, rather than using the stepping model of the typical imperative debugger. As with Hat, part of the motivation for its design choices revolves around the extra complication of laziness -- with Haskell's primitive {\small\texttt{Debug.trace}}, for example, the act of printing something out might change the evaluation order of the program, and therefore suppress a bug, or at least complicate reasoning. Hood allows the user to insert points at which observations about data structures are collected without altering the observable behaviour of the program. The authors give examples of this method fitting particularly well with the point-free style of functional programming, the observation point acting as a quasi-identity-function in the middle of a chain of functions. For example, {\small\texttt{consumer\negthinspace\ .\negthinspace\ observe\negthinspace\ "intermediate"\negthinspace\ .\negthinspace\ producer}} as the equivalent to {\small\texttt{consumer\negthinspace\ .\negthinspace\ producer}} but storing the debug information for this observation point under the label {\small\texttt{intermediate}}, from where it may be retrieved later. The Hood tool itself can be used for viewing such information. 

The 2007 GHC debugger \cite{marlow} was designed by looking at the flaws of Hat and Hood and trying to avoid them. In particular, the authors list ways in which Hat and Hood are not always available -- for example, not able to be used on all programs, or being limited to one compiler, or requiring re-compilation of libraries, or not being able to be used interactively, or not being able to print polymorphic values. They go so far as to say \textit{``The debugger should work with everything and always be available, even if this means sacrificing functionality''}. We have seen this same observation about `availability' or `accessibility' as the cornerstone of usability in our review of the literature in Section 1.1.

The debugger is used by loading the program into the REPL in the normal way, and using the extended REPL commands provided by the debugger (for example {\small\texttt{:break}}) to control debugging. Values of in-scope names may be inspected, and the program single-stepped.

An email survey \cite{fpcomplete} targeting 16000 Haskell users (with 1240 responses), commissioned by a commercial Haskell contractor, asked respondents to fill in the blank in the following sentence: \textit{``Debugging and Profiling: improvements in this would be \underline{\hphantom{4mm}}''}. The results were Crucial 29\%, Important 30\%, Helpful 23\%, Slight help 9\%, No impact 4\%. Total Crucial or Important 59\%. A free response field was also provided. The responses include \textit{``Debugging Haskell code is like groping in the dark with a hand tied behind your back.''}, \textit{``Debugging Haskell is still a pain for beginners and hampers adoption.''},
\textit{``To be honest I'm a bit `afraid' of this part of Haskell.''}, and \textit{``I would never be able to convince my coworkers \textup{[to adopt Haskell]} without decent debugging support.''} This last answer alludes to a source of past disillusionment about the apparent lack of progress of the art of programming despite vast improvements in computing power, language design, and compiler tools. As the field advances, old problems are solved only to be replaced with ones which could not have been conceived of unless we had already solved the old ones. Debugging is likely always to be needed, and unlikely to be eliminated in the way envisaged by the early pioneers of computing.

\textbf{Lisp}\hspace{3.5mm} Common Lisp \cite{commonlisp} has a tracing function similar to the OCaml one we looked at earlier, although there are more sophisticated facilities: the user can ask for certain values to be printed at each step, or for tracing to begin or end only when a certain predicate related to the code holds. The tracer is itself implemented as a LISP macro.

\smallskip
\begin{verbatim}[commandchars=\\\{\}]
[1]> (TRACE rev)                                            \textrm{\textit{trace our function}}
;; Tracing function REV.
(REV)x
[3]> (rev '(1 2 3 4))                                       \textrm{\textit{invoke it}}
1. Trace: (REV '(1 2 3 4))
2. Trace: (REV '(2 3 4))
...
3. Trace: REV ==> (4 3)
2. Trace: REV ==> (4 3 2)
1. Trace: REV ==> (4 3 2 1)
(4 3 2 1)
\end{verbatim}
\smallskip

\noindent In a similar fashion to the OCaml tracer, only the inputs and outputs are shown, rather than a diagram of the evaluation of the insides of the function.

Racket \cite{racket}, a modern Scheme implementation, contains two debugging tools. The first is  a breakpoint-based debugger with an optional graphical interface. Panes show the stack and the values of local names. When execution is paused at the start of an expression, an alternative value may be substituted for an expression, for experimentation purposes. Similarly, when execution is paused at the end of an expression's evaluation, an alternative return value may be substituted. The second is an algebraic stepper, which can show each step of the evaluation of the program in the source language:

\medskip

\noindent \includegraphics[width=0.75\textwidth]{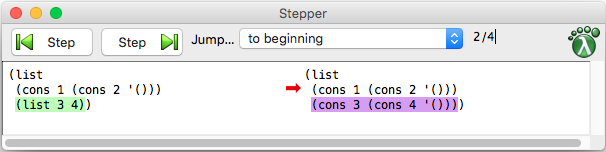}

\medskip

Both the debugger and the algebraic stepper use Racket's \textit{continuation marks} scheme \cite{Clements2001}, which elaborates the program source such that, when it is compiled, enough information remains on the stack to point to, or even reconstruct the expression at the marked points. The debugger works for all Racket programs, but does not show the actual state of the expression being evaluated. The algebraic stepper does show the actual expression, but it only works for the small ``Beginning student'' and ``Intermediate Student'' languages, not the full Racket language. A recent paper by Cong and Asai \cite{cong2016implementing} uses somewhat similar ideas to build an algebraic stepper for a small subset of Ocaml, for teaching purposes. One cannot debug into other libraries unless the libraries themselves have been compiled with such an elaboration. This may be mitigated somewhat by shipping an optional, elaborated version of the language's standard libraries. Such a  stepping approach, nevertheless, appears to offer a compelling foundation for debugging functional programs, if it could be freed from its limitations.

\textbf{Summary}\hspace{3.5mm}We have explored some of the historical context of debugging as a persistent problem, and surveyed existing solutions in the functional world. Now we shall propose an approach of our own, which is rather surprising, but which we claim might form the foundation of a way of writing debuggers for functional languages which is free of many of the disadvantages of previous solutions.

\section{Approach}

We intend to tackle the problem of debugging by directly interpreting the program, allowing the intermediate steps of evaluation to be shown. When we say ``directly interpreting'', we mean just that -- a completely naive step-by-step evaluation of the source code or AST without recourse to any kind of transformation or compilation whether involving a bytecode or not (we make this clarification because the REPL is often referred to as an interpreter though it does not work by interpretation.) Why interpretation? Because it allows the program to be run without the loss of information inherent in the compilation process. There is no reconstruction of information required, no lossy mapping back and forth between source and executable. It fits the model of functional programming as evaluation by reducing an expression to a value, rather than making the programmer think imperatively. This choice will, we hope, allow for a design which sweeps away many of the disadvantages of existing solutions, replacing them with one big disadvantage -- that interpretation is extremely slow. We shall then work to mitigate that disadvantage to arrive at a usable debugger.

Before getting too deeply into this line of thought, let us begin with an example to illustrate the concept before returning to our argument, and to the specifics of our implementation.

\subsection{An example}

\noindent As is traditional, we consider a program for calculating the factorial of a positive number:

\begin{Verbatim}[commandchars=\\\{\}]
\texttt{\textbf{let rec} factorial n =}\\
\texttt{  \textbf{if} n = 1 \textbf{then} 1 \textbf{else} n * factorial (n - 1)}\\
\texttt{\textbf{in}}\\
\texttt{  factorial 4}
\end{Verbatim}

\noindent The upper portion of Figure 1 shows a naive visualization of the evaluation of this program. This is certainly  not how we would write such an evaluation on paper. Although the evaluation shown is self-contained in the sense that each line of it is a valid program (which might seem a useful property) it is hard to see what is going on. It is large, both in width (how long the expression becomes) and length (how many lines are needed). Writing each evaluation step over multiple lines as we did with the original program above would not only increase the length, but make it difficult to visually compare adjacent lines. We must reduce the amount of information shown, even in this simple case.

Look now at the lower part of Figure 1, showing the output of our prototype system. We removed the definition of the {\small\texttt{factorial}} function itself (since it is recursive, its name will appear in the expression anyway.) We avoided printing any reduction step which leads to an expression such as {\small\texttt{\textbf{if}\negthinspace\ false}} or {\small\texttt{\textbf{if}\negthinspace\ true}}. We have not shown the intermediate steps of simple arithmetic which reduce {\small\texttt{4\negthinspace\ *\negthinspace\ (3\negthinspace\ *\negthinspace\ (2\negthinspace\ *\negthinspace\ 1))}} to {\small\texttt{24}}. We have removed trivial arithmetic (e.g. subtracting one), even when it involves variable names, such as reducing {\small\texttt{n\negthinspace\ -\negthinspace\ 1}} to {\small\texttt{3}} directly rather than via {\small\texttt{4\negthinspace\ -\negthinspace\ 1}}. We have removed {\small\texttt{\textbf{let}}} bindings which apply to the

\begin{landscape}
\scalebox{0.65}{
\begin{minipage}{2\textwidth}
{\ttfamily
~~~~let rec factorial n = if n = 1 then 1 else n * factorial (n - 1) in factorial 4\\
=>  ~let rec factorial n = if n = 1 then 1 else n * factorial (n - 1) in let n = 4 in if n = 1 then 1 else n * factorial (n - 1)\\
=>  ~let rec factorial n = if n = 1 then 1 else n * factorial (n - 1) in let n = 4 in if false then 1 else n * factorial (n - 1)\\
=>  ~let rec factorial n = if n = 1 then 1 else n * factorial (n - 1) in let n = 4 in n * factorial (n - 1)\\
=>  ~let rec factorial n = if n = 1 then 1 else n * factorial (n - 1) in let n = 4 in 4 * factorial (n - 1)\\
=>  ~let rec factorial n = if n = 1 then 1 else n * factorial (n - 1) in 4 * factorial (4 - 1)\\
=>  ~let rec factorial n = if n = 1 then 1 else n * factorial (n - 1) in 4 * factorial 3\\
=>  ~let rec factorial n = if n = 1 then 1 else n * factorial (n - 1) in 4 * (let n = 3 in if n = 1 then 1 else n * factorial (n - 1))\\
=>  ~let rec factorial n = if n = 1 then 1 else n * factorial (n - 1) in 4 * (let n = 3 in if false then 1 else n * factorial (n - 1))\\
=>  ~let rec factorial n = if n = 1 then 1 else n * factorial (n - 1) in 4 * (let n = 3 in n * factorial (n - 1))\\
=>  ~let rec factorial n = if n = 1 then 1 else n * factorial (n - 1) in 4 * (let n = 3 in 3 * factorial (n - 1))\\
=>  ~let rec factorial n = if n = 1 then 1 else n * factorial (n - 1) in 4 * (3 * factorial (3 - 1))\\
=>  ~let rec factorial n = if n = 1 then 1 else n * factorial (n - 1) in 4 * (3 * factorial 2)\\
=>  ~let rec factorial n = if n = 1 then 1 else n * factorial (n - 1) in 4 * (3 * (let n = 2 in if n = 1 then 1 else n * factorial (n - 1)))\\
=>  ~let rec factorial n = if n = 1 then 1 else n * factorial (n - 1) in 4 * (3 * (let n = 2 in if false then 1 else n * factorial (n - 1)))\\
=>  ~let rec factorial n = if n = 1 then 1 else n * factorial (n - 1) in 4 * (3 * (let n = 2 in n * factorial (n - 1)))\\
=>  ~let rec factorial n = if n = 1 then 1 else n * factorial (n - 1) in 4 * (3 * (let n = 2 in 2 * factorial (n - 1)))\\
=>  ~let rec factorial n = if n = 1 then 1 else n * factorial (n - 1) in 4 * (3 * (2 * factorial (2 - 1)))\\
=>  ~let rec factorial n = if n = 1 then 1 else n * factorial (n - 1) in 4 * (3 * (2 * factorial 1))\\
=>  ~let rec factorial n = if n = 1 then 1 else n * factorial (n - 1) in 4 * (3 * (2 * (let n = 1 in if n = 1 then 1 else n * factorial (n - 1))))\\
=>  ~let rec factorial n = if n = 1 then 1 else n * factorial (n - 1) in 4 * (3 * (2 * (let n = 1 in if true then 1 else n * factorial (n - 1))))\\
=>  ~let rec factorial n = if n = 1 then 1 else n * factorial (n - 1) in 4 * (3 * (2 * 1))\\
=>  ~let rec factorial n = if n = 1 then 1 else n * factorial (n - 1) in 4 * (3 * 2)\\
=>  ~let rec factorial n = if n = 1 then 1 else n * factorial (n - 1) in 4 * 6\\
=>  ~let rec factorial n = if n = 1 then 1 else n * factorial (n - 1) in 24\\
=>  ~24\par}
\bigskip
\bigskip
\bigskip
\bigskip
\bigskip

{\ttfamily~~~~factorial 4\\
n = 4 ~=> ~\underline{\textbf{if} n = 1 \textbf{then} 1 \textbf{else} n * factorial (n - 1)}\\
n = 4 ~=> ~n * factorial (\underline{n - 1})\\
\-~~~~~~~=> ~4 * \underline{factorial 3}\\
n = 3 ~=> ~4 * (\underline{\textbf{if} n = 1 \textbf{then} 1 \textbf{else} n * factorial (n - 1)})\\
n = 3 ~=> ~4 * (n * factorial (\underline{n - 1}))\\
\-~~~~~~~=> ~4 * (3 * \underline{factorial 2})\\
n = 2 ~=> ~4 * (3 * (\underline{\textbf{if} n = 1 \textbf{then} 1 \textbf{else} n * factorial (n - 1)}))\\
n = 2 ~=> ~4 * (3 * (n * factorial (\underline{n - 1})))\\
\-~~~~~~~=> ~4 * (3 * (2 * \underline{factorial 1}))\\
n = 1 ~=> ~4 * (3 * (2 * (\underline{\textbf{if} n = 1 \textbf{then} 1 \textbf{else} n * factorial (n - 1)})))\\
\-~~~~~~~=> ~4 * (3 * (\underline{2 * 1}))\\
\-~~~~~~~=>* 24\par}
\end{minipage}}
\bigskip
\bigskip

\noindent Figure 1. A naive rendering of the evaluation of {\small\texttt{factorial\negthinspace\ 4}} showing each step of the evaluation, followed by an automatically reduced one, eliding a) parts of the evaluation of the {\small\texttt{\textbf{if}}} construct; b) the definition of a recursive function mentioned in the expression; c) the final portion of arithmetic; and d) trivial operations such as {\small\texttt{3\negthinspace\ -\negthinspace\  1}}. In addition, {\small\texttt{\textbf{let}}} expressions unique in the whole expression are moved to the left, and basic syntax highlighting has been used. The expression to be reduced in each step has been underlined.\end{landscape}

whole expression to the left hand side of the {\small\texttt{=>}} arrow to avoid too many {\small\texttt{\textbf{let}\negthinspace\ n\negthinspace\ = \ldots}} instances making the output too wide. We have used simple syntax highlighting in the form of bold for keywords. Finally, we have underlined the expression to be reduced at each step. All changes have been made automatically. Each step is no longer a valid OCaml program, but the increase in readability is significant. Clearly, for larger programs, such elision will be even more important, since the focus needs to be on the currently-evaluating subexpression of a potentially huge expression representing the whole program. Note that all the intervening steps of the computation are performed, but certain lines are not printed. This means that the finer details of the computation may be inspected upon demand.

In the program trace we have already exhibited, it is clear that for realistic programs, the program trace (both its width and its length) may be significant. This issue is discussed in some detail by Taylor \cite{taylor-thesis} and Pajera-Flores \cite{winhipe}. A practical solution, we claim, must involve providing ways of a) eliding information within a single step -- reducing the width; b) eliding whole steps -- reducing the length; c) searching the resultant trace, if it is still too large to spot the bug; and d) moving backward and forward through the trace to connect cause and effect in the computation.

\subsection{Design choices}
\label{tests}
We choose to build a debugger which puts the notion of accessibility first (it is the core of usability), and everything else second. We claim that, without such universal availability, our debugger would be added to the growing pile of debuggers for functional languages which lie unused. And so this extremism is, in fact, in the service of practicality.

We can contrast the accessibility of a low-level debugger (such as GDB) which allows any executable to be debugged (assuming it was compiled with debug information enabled) to the much more limited accessibility of various high-level debuggers. We need to bridge this gap. So we should define precisely what trade-offs we are willing to accept. We can say that we will allow the following leeway in our extremism: 1) Just like GDB, we will insist upon the executable being compiled with a special flag. In fact, the requirement is rather stronger than GDB, since GDB is of some use on an executable not so compiled. 2) We shall not make any attempt to provide for the debugging of code from other languages (so, for example, C code linked into a primarily-OCaml executable will not be debuggable). But such executables will run properly. This means that, for example, if one suspects a bug exhibited in an OCaml executable is really a bug in the OCaml compiler or runtime, debugging it may require GDB in addition to our debugger. 3) There will doubtless be small ways in which the interpreter differs, whilst still being equivalent given the OCaml semantics. For example, its exact limits for stack overflows on non-tail-recursive code may be different, or the order of execution of threads may be different. This might mean that some bugs go away when switching to the interpreted version of OCaml. This might damage reproducibility when debugging certain kinds of low-level problems.

\noindent Notwithstanding these small things, we believe that this is the sensible approach for a high-level debugger. Let us list some of our principal aims here, to remind ourselves. We would like our eventual debugger (a) to be useable with any build system; (b) to work with mixed C/OCaml code; (c) to be able to debug libraries, not just the programmer's own code; (d) to be easy to keep in sync with the OCaml toolchain, so a new version can be released with each OCaml version; (e) to require minimal patches to the target toolchain; and (f) to be suitable for debugging the development of the OCaml compiler itself, and similar complex code.

\noindent A debugger which embodies these two kinds of requirement (negative and positive) ought to have the spirit of accessibility and so fulfil our needs.

\subsection{Correctness and Maintainence}

How will we know that the step-by-step interpreter is correct? Correctness is the major concern in any implementation of a programming language (which is why certified compilers have become such a hot topic), but we have a special extra concern. We must make sure that the interpreter matches the semantics of the native code OCaml compiler. It is no use trying to debug a program, only to find that the bug changes or disappears when using the interpreter. This concern exists also in the core OCaml distribution, unusually, since there are two compilers -- bytecode and native code.

Our interpreter shares the front end with the bytecode and native code compilers, so we need not worry about correctness there -- its behaviour should match exactly the bytecode and native code behaviour. In the actual evaluation, though, we must ensure the semantics of the language are obeyed to the same standard as in the compiled implementations. Formal proof will not work. There is no formal semantics of OCaml to prove adherence to and, in any event, without also formally proving the OCaml compiler, we cannot show the two are semantically equivalent. The OCaml distribution does come, however, with a large test suite, which we will use to test the interpreter.
How do we know that it will continue to work when the OCaml compiler is updated? Languages change, and new language features are added. Of course, the best solution would be to eventually have our small-step interpreter included in the OCaml distribution itself. The reason for doing so would be social, not technical -- it would ensure that for each release the interpreter would be updated along with the compilers. It is not unusual for debuggers to be included in the core distribution of a language. Failing such an inclusion, the debugger will have to be updated for each major release of OCaml. But these changes are likely to be rather easier than one might expect, due to the sharing of the front end. And some, such as the changes to typechecking internals in the front end which frequently appear in OCaml compiler change logs, will require no changes to the interpreter.

There is a marked difference in complexity between writing a simple big-step evaluator for an abstract syntax tree and its small-step counterpart. Recent work by Cong and Asai suggests a scheme whereby one may \textit{``implement a stepper concisely by writing an evaluator that is close to a standard big-step interpreter''} \cite{cong2016implementing}, so there is reason to believe that this difficulty may also be overcome.

\subsection{Functional program visualization}

This paper is concerned principally with the overall shape of our approach and the technical details of \textsf{OCamli}, our interpreter for OCaml. However, we must also discuss the actual mechanisms for the visualisation of programs as shown in our factorial example. There is plenty of literature on this in the field of software visualization, as well as some earlier work in the field of functional languages. We take a brief review  now (the following paragraphs appear in similar form in our earlier paper \cite{whitington}).

Two useful surveys \cite{UrquizaFuentes} \cite{survey2009} give a general overview of recent developments in this area, the first specific to functional programming, the second with wider scope. A very broad introduction \cite{PetredeQuincey} provides background. A comprehensive survey \cite{Sorva} of education systems for program visualization is useful too. We pick out a few recent systems for further discussion.

The \textit{WinHIPE} system \cite{winhipe} is a recent incarnation of these ideas for the HOPE \cite{burstall1980hope} language. It uses a step-by-step evaluation system, and explicitly addresses the problems of scale by elision of information and a focusing mechanism. The emphasis, however, is on graphical (tree-based) representations, an approach we shall not take, being of the belief that trees can often be, in fact, harder to read than well-pretty-printed program representations. The \textit{Visual Miranda Machine} \cite{visual-miranda} provides a trace of the evaluation of a lazy functional program, together with a commentary showing the reason for choosing each evaluation step. There is a discussion of granularity, taking the example of the ``list comprehension'' language feature. \textit{DrScheme} \cite{drscheme} provides, amongst many other facilities, an ``algebraic stepper'' for the Scheme language that can print out steps of evaluation. The stepper, however, supports only a subset of the language. The implementation is interesting, though -- it reuses some of the underlying Scheme implementation to ensure equivalent semantics.  Touretsky  describes a LISP-based system \cite{touretzky} that produces mainly textual traces, but with some use of graphical elements to indicate the different scoping mechanisms peculiar to LISP. The presentation of \textit{ZStep95} \cite{zstep95} begins by noting that debugging is, essentially, a human interface problem. The authors concentrate on the concept of \textit{immediacy} (temporal, spatial, and so on), which they see as essential, and exhibit a stepping debugger for a functional language which can go back and forth through time.   Another approach to this problem is as a special case of the more general concept of a \textit{calculator} \cite{Reeves95thecalculator} \cite{miracalc}, showing how it pertains to various logical systems with a mathematical basis, not just functional programs. \textit{Prospero} \cite{taylor-thesis} is a more fully-developed system, again for a lazy language. It includes methods for filtering the evaluation trace to elide information and a careful discussion of usability issues. A recent system for visualizing the evaluation of JavaScript, including its functional elements, is JSExplain \cite{chargueraud2018jsexplain}. It uses a reference interpreter derived from the language specification itself and so, in addition to its uses for teaching, it acts as a debugger for the specification of the JavaScript language itself.

These systems are mostly concerned with program visualization for teaching; we wish to bias ourselves towards the task of general debugging, hoping that some of the educational uses will be subsumed by it. The authors of DrScheme \cite{drscheme} urge caution here, choosing instead to build a ``tower'' of syntactically restrictive variants of Scheme specifically for educational purposes. They say that, due to the fact that so many sequences of characters are syntactically valid in Scheme, error messages are less confusing when the dialect is restricted -- we would prefer to avoid this in the name of universality.

It is worth pointing out that much research in software visualization concerns overtly graphical approaches. We take a simpler line, sticking to pretty-printing. We claim that the most important aspect of a successful visualization is elision -- reducing the information visible to just what is required so that large datasets may be understood easily. Programmers are used to seeing their program as text, and visualizing its evaluation as, for example, a graphical tree structure, is less useful for debugging large programs (it can be useful, of course, for visualizing program source code structure as opposed to evaluation traces).

\section{\textsf{OCamli}: an interpreter for OCaml}

This section presents the technical development and practical use of the prototype \textsf{OCamli} interpreter \cite{ocamli} for OCaml, considering both the method of step-by-step evaluation, and the heuristics used to make for concise, readable output. It does not support the whole language, but can load almost all of the OCaml Standard Library, and run its initialisation. It is the foundation upon which our eventual debugger will be built. Presently, it has a prototype command line interface which, on its own, is suitable for debugging some little OCaml programs. 

We should like to have a command {\small\texttt{ocamli}} so we may write {\small\texttt{ocamli test.ml}} and the program will be interpreted, functioning in the same way as it would if compiled and then run. We will allow the flag {\small\texttt{-show}} to show the final result of the evaluation of the program:
\smallskip
\begin{verbatim}[commandchars=|\[\]]
$ ocamli -show test.ml
7
\end{verbatim}
\smallskip

\noindent Here, the contents of {\small\texttt{test.ml}} is simply ``{\small\texttt{1 + 2 * 3}}''. The flag  {\small\texttt{-show-all}} will show all the stages of computation:

\begin{Verbatim}[commandchars=|\[\]]
$ ocamli -show-all test.ml
1 + |underline[2 * 3]
|underline[1 + 6]
7
\end{Verbatim}
\smallskip

\noindent Such a program will need to read the source code, convert it to a representation suitable for direct interpretation, interpret it in a fashion which allows for the printing of each individual step, and print those steps out in a readable way.

\subsection{A new representation for OCaml programs}

It is possible to produce a step-by-step interpreter for OCaml which operates directly upon the parse tree data type exposed by the \textsf{compiler-libs} library (which is the OCaml toolchain's library form). However, the data type is not ideal. It holds much information which is not needed for interpretation, complicating pattern matching. At each step, we must then reconstruct such extra information to ensure that it is a valid parse tree again. A very  early version of our interpreter was constructed using this method. The intent was to avoid introducing a new datatype (with maintenance issues), and to enable use of the existing OCaml prettyprinter. However, it quickly became apparent that the disadvantages outweighed the advantages. Consider, for example, the following code to add two integers, from a very early version of \textsf{OCamli}:

\smallskip
\begin{verbatim}[commandchars=\\\{\}]
| Pexp_apply (expr, args) ->
    \textbf{if} List.for_all (\textbf{fun} (_, arg) -> is_value arg) args \textbf{then}           \hfill\textrm{\textit{if all arguments are values}}
      \textbf{begin match} expr.pexp_desc \textbf{with}
      | Pexp_ident \{txt = Longident.Lident (("*" | "/" | "+" | "-") \textbf{as} op)\} -> \hfill\textrm{\textit{if an integer op}}
          \textbf{begin match} args \textbf{with}
            [(_, \{pexp_desc = Pexp_constant (Const_int a)\});
             (_, \{pexp_desc = Pexp_constant (Const_int b)\})] -> \hfill\textrm{\textit{extract integer values}}
              \textbf{let} result = calculate a b op \textbf{in}
                \{e \textbf{with} pexp_desc = Pexp_constant (Const_int result)\} \hfill\textrm{\textit{rebuild parse tree node}}
          | _ -> malformed __LOC__
          \textbf{end}
      \textbf{end}
    \textbf{else}
      \textrm{\textit{(cases where one or more arguments not yet values)}}
\end{verbatim}
\smallskip

\noindent We have to check each item in the list of things to be applied is a value, match against strings representing operators, and deal with many nested records, taking them apart and building them back up once we have evaluated a single step. Instead, we should like to be able to just write:

\smallskip
\begin{verbatim}
| Op (op, Int x, Int y) -> Int (calculate op x y)
\end{verbatim}
\smallskip

\noindent This is the aim of our new representation for OCaml programs (we will not call it a representation of OCaml parse trees or Abstract Syntax Trees, because the information it must contain is different.) We call it \textsf{TinyOCaml}. Here is part of the main type (we skip most of it for brevity, together with some of the types to which it refers). Unlike the OCaml parse tree datatype, which is a complicated set of mutually-recursive datatype definitions, we have only a few here.

\smallskip
\begin{Verbatim}[commandchars=\\\{\}]
\textbf{type} t =
  Unit                        \hfill\textrm{\textit{atomic types}}
| Int \textbf{of} int         
| Bool \textbf{of} bool       
| Float \textbf{of} float
| Record \textbf{of} (string * t ref) list \hfill\textrm{\textit{record}}
| Tuple \textbf{of} t list    \hfill\textrm{\textit{tuple}}
| Cons \textbf{of} (t * t)    \hfill\textrm{\textit{list}}
| Nil                       
| Array \textbf{of} t array   \hfill\textrm{\textit{array}}
| Constr \textbf{of} int * string * t option \hfill\textrm{\textit{user-defined data type constructor}}
| Fun \textbf{of} (label * pattern * t * env)  \hfill\textrm{\textit{function}}
| Function \textbf{of} (case list * env) \hfill\textrm{\textit{function with pattern-match}}  
| Var \textbf{of} string               \hfill\textrm{\textit{variable}}
| Op \textbf{of} (op * t * t)          \hfill\textrm{\textit{binary operator}}
| Cmp \textbf{of} (cmp * t * t)        \hfill\textrm{\textit{comparison operator}}
| If \textbf{of} (t * t * t option)    \hfill\textrm{\textit{conditional statement}}
| Let \textbf{of} (bool * binding list * t) \hfill\textrm{\textit{let-binding}}
| LetDef \textbf{of} (bool * binding list) \hfill\textrm{\textit{let-binding structure item}}
| TypeDef \textbf{of} (bool * Parsetree.type_declaration list) \hfill\textrm{\textit{user-defined type definition}}
| App \textbf{of} (t * t)              \hfill\textrm{\textit{function application}}
| Seq \textbf{of} (t * t)              \hfill\textrm{\textit{imperative \texttt{;} operator}}
| While \textbf{of} (t * t * t * t)    \hfill\textrm{\textit{while loop}}
| For \textbf{of} (string * t * forkind * t * t * t) \hfill\textrm{\textit{for loop}}
| Raise \textbf{of} (string * t option)\hfill\textrm{\textit{raise exception}}
| Match \textbf{of} (t * case list)    \hfill\textrm{\textit{pattern match}}
| TryWith \textbf{of} (t * case list)  \hfill\textrm{\textit{\texttt{\textbf{try}}\ldots\texttt{\textbf{with}}} block}
| ExceptionDef \textbf{of} (string * Parsetree.constructor_arguments) \hfill\textrm{\textit{exception definition}}
| CallBuiltIn \textbf{of} \hfill\textrm{\textit{built-in primitive}}
   (typ option * string * t list * (env -> t list -> t))
| Struct \textbf{of} (bool * t list)   \hfill\textrm{\textit{module implementation}}
| Sig \textbf{of} t list               \hfill\textrm{\textit{module signature}}
\end{Verbatim}
\smallskip

\noindent Let us look again at an example program, and see its evaluation as it may be printed on screen by \textsf{OCamli}:

\smallskip
\begin{Verbatim}[commandchars=|\[\]]
   1 + 2 > |underline[3 + 4]
=> |underline[1 + 2] > 7
=> |underline[3 > 7]
=> false
\end{Verbatim}
\smallskip

\noindent And here is what is going on inside \textsf{OCamli} -- much simpler than directly manipulating the OCaml parsetree itself:

\smallskip
\begin{Verbatim}
   Cmp (GT, Op (Add, Int 1, Int 2), Op (Add, Int 3, Int 4))
=> Cmp (GT, Op (Add, Int 1, Int 2), Int 7)
=> Cmp (GT, Int 3, Int 7)
=> Bool false 
\end{Verbatim}
\smallskip

\noindent We shall now consider how to convert the OCaml parse tree into our new type for OCaml programs.

\subsection{Conversion to and from real OCaml programs}

Consider the following extract of the {\small\texttt{of\_real\_ocaml}} reader for converting an OCaml parse tree into the \textsf{TinyOCaml} representation:

\smallskip
\begin{verbatim}
| Pexp_construct ({txt = Lident "[]"}, _) -> Nil
| Pexp_construct ({txt = Lident "::"}, Some ({pexp_desc = Pexp_tuple [e; e']})) ->
    Cons (of_real_ocaml env e, of_real_ocaml env e')
\end{verbatim}
\smallskip

\noindent This deals with the standard OCaml list syntax. Similar code deals with each other part of the OCaml syntax. Closure conversion is done at the same time (this is the {\small\texttt{env}} argument above), since it is convenient and avoids another pass.

Converting the other way (from {\textsf{TinyOCaml}} to OCaml's parse tree type) can be useful too, for example if we wish to use the built-in OCaml prettyprinter:

\smallskip
\begin{verbatim}
| Unit -> Pexp_construct ({txt = Longident.Lident "()"; loc = Location.none}, None)
| Int i -> Pexp_constant (Pconst_integer (string_of_int i, None)) 
| String s -> Pexp_constant (Pconst_string (s, None))
| Bool b ->
    Pexp_construct
      ({txt = Longident.Lident (string_of_bool b); loc = Location.none},
        None)
\end{verbatim}
\smallskip

\noindent Note again the stark difference in verbosity between our type \textsf{Tinyocaml.t} (to the left of each arrow) and OCaml's parse tree type (to the right of each arrow).

\subsection{Evaluating expressions}

For this proof-of-concept, a very simple interpreter has been produced. It has no pretensions towards performance, either by preserving space and time efficiency vis-a-vis the same program compiled and executed, or with regard to constant overheads. Its job is to provide a minimal working example for experimentation (remember our mantra: accessibility first, everything else, including speed, second).

\textbf{Evaluation strategy}\hspace{3.5mm} To evaluate a step of a program (that is, something of type \textsf{Tinyocaml.t}), we must first determine if the program is a value. If it is, there is no evaluation to be done. If not, we find the reducible expression, following the OCaml order of evaluation (to the extent that OCaml specifies the order). We perform one step of evaluation only. This new expression may now be returned for printing, and we continue with the next step.

Let us look at a simple example, comparing with a traditional evaluator (whose job is to evaluate down to a value in one continuous operation). We choose the short-circuiting boolean conjunction operator {\small\texttt{\&\&}}. Here is a snippet from an all-at-once interpreter: 

\smallskip
\begin{Verbatim}[commandchars=\\\{\}]
\textbf{let rec} eval = \textbf{function}
  | And (a, b) ->
      \textbf{match} eval a \textbf{with}
      | Bool false -> Bool false
      | Bool true -> eval b
\end{Verbatim}
\smallskip

\noindent We evaluate the left hand side {\small\texttt{a}} to a \textsf{Tinyocaml.t} representing a boolean (either {\small\texttt{Bool\negthinspace\ true}} or {\small\texttt{Bool\negthinspace\ false}}). If it is {\small\texttt{Bool\negthinspace\ false}}, this is the result. If it is {\small\texttt{Bool\negthinspace\ true}}, the code evaluates the right hand side to a value, and returns it.
 Contrast with the following, which evaluates just a single step:

\smallskip
\begin{Verbatim}[commandchars=\\\{\}]
\textbf{let rec} eval_step = \textbf{function}
  | And (Bool false, _) -> Bool false
  | And (Bool true, Bool b) -> Bool b
  | And (Bool true, b) -> eval_step b
  | And (a, b) -> And (eval_step a, b)
\end{Verbatim}
\smallskip

\noindent The first line of the pattern match in the step-by-step example is used when the left hand side has already been fully evaluated and is false: this is the short circuit. The second deals with a fully-evaluated true left hand side, and a fully evaluated right hand side. The third is the same as the second, but for a right hand side not yet fully evaluated: we have found the step which requires evaluation. The fourth and last is for an unevaluated left hand side: we evaluate the left hand side one step and leave the right hand side alone. By similar mechanisms it is possible to write a step-by-step evaluator for each other part of the language. We consider some of the more interesting ones now by way of further example.

\textbf{Imperative programs}\hspace{3.5mm} Whilst OCaml is a functional language first, there is occasional use of imperative features, and we need to display them in a way which fits in. Consider the evaluation of the OCaml {\small\textbf{\texttt{for}}} construct. When compiled, the following piece of code will print {\small\texttt{12345}}:

\smallskip
\begin{Verbatim}[commandchars=\\\{\}]
\textbf{for} y = 0 + 1 \textbf{to} 6 - 1 \textbf{do} print_int y \textbf{done}
\end{Verbatim}
\smallskip

\noindent But how do we show it? It can be treated as an expression:

\smallskip
\begin{Verbatim}[commandchars=|\[\]]
   |textbf[for] y = |underline[0 + 1] |textbf[to] 6 - 1 |textbf[do] print_int y |textbf[done]
=> |textbf[for] y = 1 |textbf[to] |underline[6 - 1] |textbf[do] print_int y |textbf[done]
=> |underline[|textbf[for] y = 1 |textbf[to] 5 |textbf[do] print_int y |textbf[done]]
1=> |underline[|textbf[for] y = 2 |textbf[to] 5 |textbf[do] print_int y |textbf[done]]
2=> |underline[|textbf[for] y = 3 |textbf[to] 5 |textbf[do] print_int y |textbf[done]]
3=> |underline[|textbf[for] y = 4 |textbf[to] 5 |textbf[do] print_int y |textbf[done]]
4=> |underline[|textbf[for] y = 5 |textbf[to] 5 |textbf[do] print_int y |textbf[done]]
5=> |underline[|textbf[for] y = 6 |textbf[to] 5 |textbf[do] print_int y |textbf[done]]
=> ()
\end{Verbatim}
\smallskip

\noindent Helpfully, the semantics of OCaml are such that {\small\texttt{\textbf{for}\negthinspace\ y\negthinspace\ =\negthinspace\ 6\negthinspace\ \textbf{to}\negthinspace\ 5\negthinspace\ \textbf{do}\negthinspace\ \ldots\negthinspace\ \textbf{done}}} is legal and does not execute the body, so we have a proper terminating condition. How is this implemented? The {\small\texttt{For}} constructor of the \textsf{Tinyocaml.t} data type looks like this:

\smallskip
\begin{Verbatim}[commandchars=\\\{\}]
For \textbf{of} string * t * forkind * t * t * t
\end{Verbatim}
\smallskip

\noindent Our example would be represented like this:

\smallskip
\begin{Verbatim}
  For ("y",
       Op (Add, Int 0, Int 1),
       UpTo,
       Op (Sub, Int 6, Int 1),
       App (Var "print_int", Var "y"),
       App (Var "print_int", Var "y"))
\end{Verbatim}  
\smallskip

\noindent We need two copies of the body, so that one may be evaluated step-by-step, and then, when it has been reduced to a value, the spare copy can be moved into place, and we go round again. Here are all the cases needed for step-by-step evaluation of the {\small\textbf{\texttt{for}}} construct:

\smallskip
\begin{Verbatim}[commandchars=\\\{\}]
| For (v, e, ud, e', e'', copy) \textbf{when} not (is_value e) ->        \hfill\textrm{\textit{evaluate from part}}
    For (v, eval env e, ud, e', e'', copy)
| For (v, e, ud, e', e'', copy) \textbf{when} not (is_value e') ->       \hfill\textrm{\textit{evaluate to part}}
    For (v, e, ud, eval env e', e'', copy)
| For (_, Int x, UpTo, Int y, _, _) \textbf{when} x > y -> Unit          \hfill\textrm{\textit{end condition}}
| For (_, Int x, DownTo, Int y, _, _) \textbf{when} y > x -> Unit        \hfill\textrm{\textit{end condition}}
| For (v, Int x, ud, e', e'', copy) \textbf{when} is_value e'' ->        \hfill\textrm{\textit{advance the \textbf{\texttt{for}} loop using the copy}}
    For (v, Int (x + 1), ud, e', copy, copy)
| For (v, x, ud, e', e'', copy) ->                                       \hfill\textrm{\textit{evaluate the body}}
    For (v, x, ud, e', eval (EnvBinding (false, ref [(PatVar v, x)])::env) e'', copy)
\end{Verbatim}
\smallskip

\noindent Note the final case, where the variable is bound for the evaluation. The treatment of {\small\texttt{\textbf{while}}} is similar. Now consider how to deal with another imperative construct: the reference. A reference in OCaml is a mutable cell containing a value. Here is a simple imperative program using a reference:

\smallskip
\begin{verbatim}[commandchars=\\\{\}]
    \textbf{let} x = \underline{ref 0} \textbf{in} x := !x + 1
=>  \textbf{let} x = \{contents = 0\} \textbf{in} x := \underline{!x} + 1
=>  \textbf{let} x = \{contents = 0\} \textbf{in} x := \underline{0 + 1}
=>  \textbf{let} x = \{contents = 0\} \textbf{in} \underline{x := 1}
=>  \underline{\textbf{let} x = \{contents = 1\} \textbf{in} ()}
=>  ()
\end{verbatim}
\smallskip

\noindent Note that, even though the new value of the reference is lost in the final expression {\small\texttt{()}}, it is visible in the penultimate step, which is good enough. \textsf{OCamli} can emulate the low-level primitives used to implement some of OCaml's basic language features. When we opt to show the low-level primitives involved in the use of references, we see a somewhat longer version:

\smallskip
\begin{verbatim}[commandchars=\\\{\}]
    \textbf{let} x = \underline{ref 0} \textbf{in} x := (!x + 1)
=>  \textbf{let} x = \textbf{let} x = 0 \textbf{in} \underline{<<%makemutable x>>} \textbf{in} x := (!x + 1)
=>  \textbf{let} x = \{contents = 0\} \textbf{in} \underline{x} := (!x + 1)
=>  \textbf{let} x = \{contents = 0\} \textbf{in} \underline{\{contents = 0\} := (!x + 1)}
=>  \textbf{let} x = \{contents = 0\} \textbf{in} \underline{(\textbf{let} x = \{contents = 0\} \textbf{in fun} y -> <<%setfield0 x y>>)} (!x + 1)
=>  (\textbf{fun} y -> \textbf{let} x = \{contents = 0\} \textbf{in} <<%setfield0 x y>>) (\underline{!\{contents = 0\}} + 1)
=>  (\textbf{fun} y -> \textbf{let} x = \{contents = 0\} \textbf{in} <<%setfield0 x y>>)
    ((\textbf{let} x = \{contents = 0\} \textbf{in} \underline{<<%field0 x>>}) + 1)
=>  (\textbf{fun} y -> \textbf{let} x = \{contents = 0\} \textbf{in} <<%setfield0 x y>>) \underline{(0 + 1)}
=>  \underline{(\textbf{fun} y -> \textbf{let} x = \{contents = 0\} \textbf{in} <<%setfield0 x y>>)} 1
=>  \textbf{let} y = 1 \textbf{in let} x = \{contents = 0\} \textbf{in} \underline{<<%setfield0 x y>>}
=>  ()
\end{verbatim}
\smallskip

\noindent Most users will not want this longer output by default, but it is helpful when we wish to see, for example, exactly what I/O calls are triggered by Standard Library functions.

\textbf{Currying}\hspace{3.5mm} When we teach functional programming we often say ``every function only has one argument'' but really, except in cases of partial application, programmers think of curried functions as a single function of multiple arguments. And how the programmer thinks is how the debugger must behave. Consider the default evaluation of {\small\texttt{(\textbf{fun}\negthinspace\ x\negthinspace\ y\negthinspace\ ->\negthinspace\ x\negthinspace\ +\negthinspace\ y)\negthinspace\ 4\negthinspace\ 5}}:

\smallskip
\begin{verbatim}[commandchars=\\\{\}]
    \underline{(\textbf{fun} x y -> x + y) 4} 5
=>  \underline{(\textbf{let} x = 4 \textbf{in fun} y -> x + y)} 5
=>  \underline{(\textbf{fun} y -> \textbf{let} x = 4 \textbf{in} x + y) 5}
=>  \textbf{let} y = 5 \textbf{in let} x = 4 \textbf{in} \underline{x + y}
=>  \textbf{let} y = 5 \textbf{in} \underline{4 + y}
=>  \underline{4 + 5}
=>  9
\end{verbatim}
\smallskip

\noindent We always print {\small\texttt{\textbf{fun}\negthinspace\ x\negthinspace\ y\negthinspace\ ->}} instead of {\small\texttt{\textbf{fun}\negthinspace\ x\negthinspace\ ->\negthinspace\ \textbf{fun}\negthinspace\ y\negthinspace\ ->}}, since they are indistinguishable in the OCaml parse tree. There are a small number of places where semantically equivalent syntactic forms are not distinguished like this, and we would want eventually to modify the OCaml parser to retain information about the original form.

Returning to currying, the evaluation above is excessively verbose. When the {\small\texttt{-fast-curry}} option is added to the command line, the arguments will be applied at once:

\smallskip
\begin{verbatim}[commandchars=\\\{\}]
    \underline{(\textbf{fun} x y -> x + y) 4 5}
=>  \textbf{let} x = 4 \textbf{in let} y = 5 \textbf{in} \underline{x + y}
=>  \textbf{let} y = 5 \textbf{in} \underline{4 + y}
=>  \underline{4 + 5}
=>  9
\end{verbatim}
\smallskip

\noindent This involves a more complicated matching on the program to identify all the arguments which can be applied. It is an example of the requirements of the visualization driving the implementation of the interpreter's evaluation model. In fact, combined with another option {\small\texttt{-side-lets}} (which pulls out let-bindings to the side), we get an evaluation which is better still:

\smallskip
\begin{verbatim}[commandchars=\\\{\}]
    \underline{(\textbf{fun} x y -> x + y) 4 5}
x = 4 y = 5 =>  \underline{x + y}
      y = 5 =>  \underline{4 + y}
            =>  \underline{4 + 5}
            =>  9
\end{verbatim}
\smallskip

\noindent In the future, we could go further, and do away with the step-by-step lookup of variables, imagining the optimal visualisation:

\smallskip
\begin{verbatim}[commandchars=\\\{\}]
    \underline{(\textbf{fun} x y -> x + y) 4 5}
x = 4 y = 5 => \underline{x + y}
            => \underline{4 + 5}
            => 9
\end{verbatim}
\smallskip

\noindent or even:

\smallskip
\begin{verbatim}[commandchars=\\\{\}]
    \underline{(\textbf{fun} x y -> x + y) 4 5}
=>  \underline{4 + 5}
=>  9
\end{verbatim}
\smallskip

\noindent This is perhaps what we might write if we were to do this on paper -- when we write such evaluations informally we naturally skip ``obvious'' steps. It is the same when doing mathematics. We can see that most of the job of improving upon the naive visualization consists of removing information, rather than adding it.

\textbf{Exceptions}\hspace{3.5mm} As one would expect, exceptions (which interrupt the flow of evaluation) require a similar mechanism inside an interpreter. The complication of a step-by-step interpreter is that exceptions must be modelled in a step-by-step way too: we cannot let uncaught raises cascade all at once. The solution to this is to model exceptions in two ways: as a special \textsf{Tinyocaml.t} constructor {\small\texttt{Raise}} and using actual exceptions. Here is the exception definition we will use, which represents, for example, the result of evaluating {\small\texttt{\textbf{raise}\negthinspace\ (Failure\negthinspace\ "broken")}} as {\small\texttt{ExceptionRaised\negthinspace\ ("Failure",\negthinspace\ Some\negthinspace\ (String\negthinspace\ "broken"))}}:

\smallskip
\begin{verbatim}[commandchars=\\\{\}]
\textbf{exception} ExceptionRaised \textbf{of} string * Tinyocaml.t option
\end{verbatim}
\smallskip

\noindent Here is the \textsf{Tinyocaml.t} constructor used to represent exceptions which need to be raised:

\smallskip
\begin{verbatim}[commandchars=\\\{\}]
| Raise \textbf{of} string * Tinyocaml.t option
\end{verbatim}
\smallskip

\noindent See how it mirrors the exception definition above. Now, let us consider the case of dividing two numbers, where the second may be zero. Here is the code from the interpreter:

\smallskip
\begin{verbatim}[commandchars=\\\{\}]
| Op (op, Int a, Int b) ->
    \textbf{begin try} Int (calc op a b) \textbf{with}
      Division_by_zero -> Raise ("Division_by_zero", None)
    \textbf{end}
\end{verbatim}
\smallskip

\noindent We use OCaml exception handling to check for {\small\texttt{Division\_by\_zero}} in the {\small\texttt{calc}} function, and if we see it, we build the {\small\texttt{Raise}} constructor as the result of evaluating this expression one step. This freezes the exception. What happens when, in the next step of evaluation, this {\small\texttt{Raise}} is found?

\smallskip
\begin{verbatim}[commandchars=\\\{\}]
| Raise (e, payload) -> \hfill\textrm{\textit{\texttt{payload} is the data carried with the exception}}
    \textbf{match} payload \textbf{with}
    | Some x \textbf{when} not (is_value x) ->
        Raise (e, Some (eval_step env x)) \hfill\textrm{\textit{if payload not a value, evaluate one step}}
    | _ ->
      \textbf{raise} (ExceptionRaised (e, payload)) \hfill\textrm{\textit{otherwise, the exception may be raised}}
\end{verbatim}
\smallskip

\noindent We may need to evaluate the expression in the {\small\texttt{Raise}} one step if it is not a value (it might be {\small\texttt{\textbf{raise} (Fail (1\negthinspace\ +\negthinspace\ 2))}}, for example). Thus, the {\small\texttt{Raise}} may take several steps to be processed. If it is a value, though, we can raise the actual exception. This will be caught in the evaluator, and mirrors the effect of the exception occurring in a compiled program. Here is code for the {\small\texttt{\textbf{try}\negthinspace\ \ldots\negthinspace\ \textbf{with}}} construct:

\smallskip
\begin{verbatim}[commandchars=\\\{\}]
| TryWith (e, cases) ->
    \textbf{if} is_value e \textbf{then} e \textbf{else} \hfill\textrm{\textit{if body a value, return}}
      \textbf{begin try} TryWith (eval_step env e, cases) \textbf{with} \hfill\textrm{\textit{evaluate body one step}}
        ExceptionRaised (x, payload) -> \hfill\textit{\textrm{if this step caused an exception}}
          \textbf{match} eval_match_exception env x payload cases \textbf{with} \hfill\textit{\textrm{see if it matches a case}}
          | FailedToMatch -> Raise (x, payload) \hfill\textit{\textrm{if not, recreate the raise node}}
          | Matched e' -> e' \hfill\textit{\textrm{otherwise, return the body of the matched case}}
      \textbf{end}
\end{verbatim}
\smallskip

\noindent If the exception is not surrounded by a {\small\texttt{\textbf{try}\negthinspace\ \ldots\negthinspace\ \textbf{with}}}, it is not caught, and so is printed at the top level and the interpreter exits:

\smallskip
\begin{verbatim}[commandchars=\\\{\}]
    1 + 1 / (\underline{1 - 1})
=>  1 + \underline{1 / 0}
=>  1 + \underline{\textbf{raise} Division_by_zero}
Exception: Division_by_zero.
\end{verbatim}
\smallskip

\noindent Let us add a {\small\texttt{\textbf{try}\negthinspace\ \ldots\negthinspace\ \textbf{with}}}:

\smallskip
\begin{verbatim}[commandchars=\\\{\}]
    \textbf{try} 1 + 1 / \underline{(1 - 1}) \textbf{with} Division_by_zero -> 2 + 2 
=>  \textbf{try} 1 + \underline{1 / 0} \textbf{with} Division_by_zero -> 2 + 2 
=>  \textbf{try} 1 + \underline{\textbf{raise} Division_by_zero} \textbf{with} Division_by_zero -> 2 + 2 
=>  \underline{2 + 2}
=>  4
\end{verbatim}
\smallskip

\noindent Now we can see the whole process. As an improvement, we might like to annotate the penultimate step to indicate which expression matched.

\textbf{Pattern matching}\hspace{3.5mm} How can we visualize pattern matching, one of the most widely-used, and praised, features of functional programming? Do we show the whole pattern, then jump to the right hand side of the chosen match? Do we show how the match matches? Consider an example:

\smallskip
\begin{verbatim}[commandchars=\\\{\}]
    \textbf{match} \underline{1 + 2} \textbf{with} 4 -> 0 | 3 -> 1 + 2 | _ -> 1 
=>  \underline{\textbf{match} 3 \textbf{with} 4 -> 0 | 3 -> 1 + 2 | _ -> 1} 
=>  \underline{\textbf{match} 3 \textbf{with} 3 -> 1 + 2 | _ -> 1} 
=>  \underline{1 + 2}
=>  3
\end{verbatim}
\smallskip

\noindent In this method of visualisation, we simply show the whole match expression with all its cases, and each time a case does not match, we drop it from the front.  As an option, in the future, we will allow the skipping of  this process, and show just the case that matched. Functional programmers are good at spotting which case will be taken, and identifying when an unexpected one has been taken, signifying a bug.

\textbf{Summary}\hspace{3.5mm} We now have a function which, given a program, can evaluate it one step. By calling the function repeatedly, feeding its own output back in as the next input, the program can be evaluated completely, step-by-step. With an appropriate pretty-printer, each step may be printed out. Sometimes the requirements of good visualization force changes to the evaluation method itself, changing the number or kind of steps.

\subsection{Dealing with size by elision}

This section concerns the important task of making the output readable (we discuss searching, which also reduces the output, in Section \ref{searching}). So, what remains? Three things: 1) the showing or eliding of whole steps  for things like simple arithmetic and variable lookups;  2) the hiding or showing of parts of the expression at each step; and 3) the default heuristics for eliding parts of individual expressions (for example, the internals of built-in functions).

For our first example, we shall consider how to automatically abridge the following arithmetic evaluation, of a type which frequently occurs at the end of a non-tail-recursive function application:

\smallskip
\begin{verbatim}[commandchars=\\\{\}]
   1 * (2 * (\underline{3 * 4}))
=> 1 * (\underline{2 * 12})
=> \underline{1 * 24}
=> 24 
\end{verbatim}
\smallskip

\noindent We wish to remove the middle two steps, leaving just:

\smallskip
\begin{verbatim}[commandchars=\\\{\}]
   \underline{1 * (2 * (3 * 4))}
=> 24
\end{verbatim}
\smallskip

\noindent This can be done by a mechanism we call peeking.

\textbf{Peeking}\hspace{3.5mm}In order to decide whether to show the current state, it is sometimes important to know the next state, and to remember the previous state. But how can we know the next state without evaluating it? One way, of course, would be to evaluate the whole program and print out its steps of execution offline. But we may wish to stop evaluation based on what is about to happen, and we cannot do this with a real running program with side effects, since we cannot roll back a side effect such as a network communication with a third party.

The solution is to add to the step-by-step evaluator the notion of \textit{peeking}. In this mode, the evaluator identifies the reducible expression, but does not evaluate it. The calling function can then interrogate the interpreter to ask ``What kind of operation would have been performed?''. Presently, the answer is one of a short list, giving just enough information to provide for some of the elisions the \textsf{OCamli} prototype can perform:

\smallskip
\begin{verbatim}[commandchars=\\\{\}]
\textbf{type} last_op =
    Arith                   \textrm{\textit{simple arithmetic}}
  | Boolean                 \textrm{\textit{\texttt{\&\&}, \texttt{||}}}
  | Comparison              \textrm{\textit{comparison operators}}
  | IfBool                  \textrm{\textit{\texttt{\textbf{if} true}, \texttt{\textbf{if} false}}}
  | InsideBuiltIn           \textrm{\textit{evaluation inside an external piece of code}}
  | VarLookup               \textrm{\textit{variable lookup}}
\end{verbatim}
\smallskip

\noindent In our example, we print the step if and only if a) the next state is a value or b) the current state is a value or c) {\small\texttt{Arith}} is not present for the previous state or d) {\small\texttt{Arith}} is not present for the next state. These four conditions, taken together, elide just enough steps of the arithmetic, but do not remove information we want to see. Similar conditions have been devised for the other kinds of elision listed in the {\small\texttt{last\_op}} type.

\textbf{Eliding within a step}\hspace{3.5mm} Consider the following example with multiple structure items (a structure item in the parlance of the OCaml parse tree is a type definition or a top level let-binding):

\begin{verbatim}[commandchars=\\\{\}]
\textbf{let} x = 1 + 2

\textbf{let} y = x + x

\textbf{let} z = 1 + y
\end{verbatim}

\noindent First, of course, we begin by evaluating {\small\texttt{1 + 2}}, and proceed from there. However, a lot of screen space is used by printing out these five lines (three code, two blank) for each step, and it can be hard for the user to follow along. Should we remove a structure item when it is no longer needed, assuming that the user is interested only in the final result of {\small\texttt{z}}? This results in a shorter but arguably incomplete trace. Or, instead, only show the structure item which is currently being evaluated? Most likely, this would be a configurable option with a sensible default, which is probably to reduce the trace as much as possible.

\subsection{The Standard Library}
OCaml comes with a small but useful library of routines. These fall broadly into three categories: (a) those which are simply there to provide a selection of common routines, useful for many programs, but which the user could write themselves -- entirely in OCaml -- if they wanted. For example, {\small\texttt{List.map}}; (b) those which are in the Standard Library because they are used in the implementation of the OCaml toolchain, but seemed to the authors to be generic enough as to be useful for the general programmer (when a programming language is in its infancy, the general programmer and the compiler author are one and the same); and (c) those which must be in the Standard Library because they provide facilities which pure OCaml programs could not provide, or use an external symbol, or talk to the runtime.

\noindent Categories (a) and (b) are easy to deal with -- we are just interpreting standard OCaml code, so it is as if the user had themselves supplied the code. The \textsf{OCamli} interpreter knows how to load multiple modules as libraries using command line arguments. For example, the following command line loads modules {\small\texttt{A}} and {\small\texttt{B}}, performing any module initialisation code, then executes the code given in the {\small\texttt{-e}} argument in an environment in which such modules exist:

\smallskip
\begin{verbatim}
$ ocamli a.ml b.ml -e 'let () = B.calc 10'
\end{verbatim}
\smallskip

\noindent It is category (c) which requires special treatment. Functions which are external to OCaml are introduced like this:

\smallskip
\begin{verbatim}[commandchars=\\\{\}]
\textbf{external} word_size : unit -> int = "%word_size"
\end{verbatim}
\smallskip

\noindent This name might be exported directly or might be used in the definition of a Standard Library function which is then exported. In the example above, it indicates that a function of type {\small\textsf{\textbf{unit} $\rightarrow$ \textbf{int}}} is expected to be available at link time under the symbol {\small\texttt{\%word\_size}} and that it is to be given the name {\small\texttt{word_size}}. When we come across such an {\small\texttt{\textbf{external}}} declaration in a {\small\texttt{.ml}} file (such as when loading the Standard Library), how should \textsf{OCamli} deal with it? What we do is to write (or generate) a binding for it. The \textsf{Tinyocaml.t} datatype already exhibited contains the constructor {\small\texttt{CallBuiltIn}}:

\smallskip
\begin{verbatim}[commandchars=\\\{\}]
CallBuiltIn \textbf{of} \ldots\ * (Tinyocaml.env -> Tinyocaml.t list -> Tinyocaml.t)
\end{verbatim}
\smallskip

\noindent This inclusion of a native OCaml function into the \textsf{Tinyocaml.t} data type for programs is the mechanism by which the gap between the interpreted and native worlds is bridged. It represents an OCaml function which takes an environment and a list of \textsf{Tinyocaml.t} arguments, calls some external native function and returns a \textsf{Tinyocaml.t} result.

We can use this {\small\texttt{CallBuiltIn}} mechanism to build an interface to our function, and a way to look up such an interface by name so that, at runtime, it may be located and called by the interpreter:

\smallskip
\begin{verbatim}[commandchars=\\\{\}]
\textbf{external} word_size : unit -> int = "%word_size"

\textbf{let} percent_word_size =
  \textbf{let} f =
    (\textbf{function} [Unit] ->
      \textbf{begin try} Int (word_size ()) \textbf{with} e -> exception_from_ocaml e \textbf{end}
     | _ -> failwith "%word_size")
  \textbf{in}
    ("%word_size",
     Fun (NoLabel, PatVar "*x", CallBuiltIn (None, "%word_size", [Var "*x"], f), []))
\end{verbatim}
\smallskip

\noindent Notice the {\small\texttt{\textbf{external}}} declaration is retained. We then create an entry to look up this function in the table of primitives {\small\texttt{("\%word\_size",\negthinspace\ x)}} where {\small\texttt{x}} is a function containing a {\small\texttt{CallBuiltIn}}. This table will be used for lookup when an {\small\texttt{\textbf{external}}} declaration is found in a {\small\texttt{.ml}} source file being interpreted. In this case, it is a function of one argument {\small\texttt{*x}} (the asterisk is a crude mechanism to mark such functions so they are not printed, since they are not part of the original source code). This function can then be applied to an argument in the interpreted world. The argument will be assigned the name {\small\texttt{*x}} and used by the native function -- the result will be returned to the interpreted world. Now we need to look at the function {\small\texttt{f}} itself. It pattern-matches on the input argument list, requiring just one argument {\small\texttt{[Unit]}}. It tries to produce the output {\small\texttt{Int\negthinspace\ (word\_size\negthinspace\  ())}} by applying the native function {\small\texttt{word\_size}} as defined by the {\small\texttt{\textbf{external}}}. This is the result. Should an exception be raised during the execution of {\small\texttt{word\_size}} (either in OCaml code or C code) it comes into the OCaml runtime as an OCaml exception, and is then converted into a \textsf{Tinyocaml.t} representation of an exception by the function {\small\texttt{exception\_from\_ocaml}}. Curried functions may be defined using a helper for each arity. For example, for arity three:

\smallskip
\begin{verbatim}[commandchars=\\\{\}]
\textbf{let} mk3 name f =
  (name,
   Fun (NoLabel, PatVar "*x",
     Fun (NoLabel, PatVar "*y",
       Fun (NoLabel, PatVar "*z",
            CallBuiltIn (None, name, [Var "*x"; Var "*y"; Var "*z"], f), []), []), []))
\end{verbatim}
\smallskip

\noindent A function defined by this method may be partially applied as usual: only when all the arguments are actually applied in the interpreter will the native function {\small\texttt{f}} be run. To avoid writing all these bindings for the Standard Library by hand, a system has been developed which allows one to write, instead:

\begin{verbatim}[commandchars=\\\{\}]
[%%auto \textbf{external} string_of_float : float -> string = "%string_of_float"]
\end{verbatim}

\noindent The binding is then generated automatically. This system, works for most of the Standard Library functions, and so reduces \textsf{OCamli}'s Standard Library file to a third of its previous size. Thus, we keep the part of \textsf{OCamli} which may need updating when OCaml is updated as small as possible. In Section 3.7, we shall see such a system might be expanded to allow an interface between compiled and interpreted code, and identify its limitations.

\subsection{Searching}
\label{searching}

We have discussed various mechanisms for making sure that \textsf{OCamli}'s output is reasonable in the default case, and that there are options for deciding what information to display. But we will want a proper searching mechanism too, especially for interactive scenarios. Of course, one way is to use standard command line tools like {\small\texttt{grep}}. How well would that work for a typical search on a typical program? We can foresee problems -- for example, patterns may need to match independent of parenthesisation. In essence, we are searching the text not the program's syntactic structure.

The problem of searching in program code, either in textual or AST form, is known in the literature. Paul and Prakash's SCRUPLE system \cite{paul1994framework} uses an extended form of the programming language's own grammar, an approach from which we shall draw inspiration. Devanbu's GENOA \cite{devanbu1999genoa} also reuses the language's parser in the context of source code analysis. Crew's ASTLOG \cite{crew1997astlog} has similar aims. The distinction between ``lexical matchers'' (such as regular expressions) and ``syntactic matchers'' (which know the syntactic structure of what they are searching) is explored in Griswold et al's TAWK system \cite{griswold1996fast}.

If we are to provide tools of our own, what facilities might be useful? Here are the basic options provided in \textsf{OCamli}:

\begin{verbatim}[commandchars=\\\{\}]
-search                  \textrm{\textit{show only matching evaluation steps}}
-highlight               \textrm{\textit{highlight the matching part of each matched step}}
-no-parens               \textrm{\textit{ignore parentheses when matching}}
-regexp                  \textrm{\textit{search terms are regular expressions rather than the built-in system}}
-upto <n>                \textrm{\textit{show the \textup{n} lines preceding each result line}}
\end{verbatim}

\noindent For example, consider searching only for lines containing {\small\verb!'4::'!}:

\smallskip
\begin{verbatim}[commandchars=\\\{\}]
=>  2::3::\textbf{let} l = [] \textbf{in let} f x = x + 1 \textbf{in} 4::map \underline{f} l
=>  2::3::\textbf{let} l = [] \textbf{in} 4::\underline{map (\textbf{fun} x -> x + 1)} l
=>  2::3::\textbf{let} l = [] \textbf{in}
      4::(\underline{\textbf{let} f x = x + 1 \textbf{in function} [] -> [] | a::l -> \textbf{let} r = f a \textbf{in} r::map f l}) l
=>  2::3::\textbf{let} l = [] \textbf{in} 
      4::(\underline{\textbf{function} [] -> [] | a::l -> \textbf{let} f x = x + 1 \textbf{in let} r = f a \textbf{in} r::map f l}) l
=>  2::3::\textbf{let} l = [] \textbf{in}
      4::(\textbf{function} [] -> [] | a::l -> \textbf{let} f x = x + 1 \textbf{in let} r = f a \textbf{in} r::map f l ) \underline{l}
=>  2::3::4::\underline{(\textbf{function} [] -> [] | a::l -> \textbf{let} f x = x + 1 in \textbf{let} r = f a \textbf{in} r::map f l ) []}
\end{verbatim}
\smallskip

\noindent This shows only the evaluation steps containing the text ``{\small\texttt{4::}}'', that is the ones where the list has almost been processed. Our search syntax is tailored to the job of searching \textsf{OCamli}'s output. The search pattern is parsed using OCaml's lexer, and then we allow any amount of whitespace between tokens, skip parentheses (if {\small\texttt{-no-parens}} is set), and allow the underscore character {\small\texttt{\_}} to stand for any token. A regular expression is generated to represent this, and searching proceeds. For example, we can search with {\small\verb!-search '[_; _; _]'!} for only those steps of evaluation which contain lists of length exactly three:

\smallskip
\begin{verbatim}[commandchars=\\\{\}]
    \underline{List.map (\textbf{fun} x -> x + 1)} [1; 2; 3]
=>  \underline{(\textbf{let} f x = x + 1 \textbf{in function} [] -> [] | a::l -> \textbf{let} r = f a \textbf{in} r::map f l )} [1; 2; 3]
=>  \underline{(\textbf{function} [] -> [] | a::l -> \textbf{let} f x = x + 1 \textbf{in let} r = f a \textbf{in} r::map f l )} [1; 2; 3]
=>  \underline{(\textbf{function} a::l -> \textbf{let} f x = x + 1 \textbf{in let} r = f a \textbf{in} r::map f l ) [1; 2; 3]}
=>  [2; 3; 4]
\end{verbatim}
\smallskip

\noindent The searches may be highlighted with {\small\texttt{-highlight}}:

\smallskip

{\footnotesize\noindent$\texttt{~~~~\underline{List.map (\textbf{fun} x -> x + 1)} \colorbox{black}{\color{white}[1; 2; 3]}}$\\
$\texttt{=>~~\underline{(\textbf{let} f x = x + 1 \textbf{in} \textbf{function} [] -> [] | a::l -> \textbf{let} r = f a \textbf{in} r::map f l )} \colorbox{black}{\color{white}[1; 2; 3]}}$\\
$\texttt{=>~~\underline{(\textbf{function} [] -> [] | a::l -> \textbf{let} f x = x + 1 \textbf{in let} r = f a \textbf{in} r::map f l )} \colorbox{black}{\color{white}[1; 2; 3]}}$\\
$\texttt{=>~~\underline{(\textbf{function} a::l -> \textbf{let} f x = x + 1 \textbf{in let} r = f a \textbf{in} r::map f l ) \colorbox{black}{\color{white}[1; 2; 3]}}}$\\
$\texttt{=>~~\colorbox{black}{\color{white}[2; 3; 4]}}$}

\smallskip

\noindent There are also options to alter the type and number of results:

\smallskip
\begin{verbatim}[commandchars=\\\{\}]
-invert-search            \textrm{\textit{invert the search, showing non-matching steps}}
-n                        \textrm{\textit{show only n results}}
-until                    \textrm{\textit{show only until this matches a printed step}}
-after                    \textrm{\textit{show only after this matches a printed step}}
-until-any                \textrm{\textit{show only until this matches any step}}
-after-any                \textrm{\textit{show only after this matches any step}}
-invert-after             \textrm{\textit{invert the after condition}}
-invert-until             \textrm{\textit{invert the until condition}}
-stop                     \textrm{\textit{stop computation after final search results}}
-repeat                   \textrm{\textit{allow the after \ldots until result to be repeated}}
\end{verbatim}
\smallskip

\noindent These options allow the programmer to show output only after a search matches, and only until another search matches. For example with {\small\verb!-after '3 + 1' -until '2::3::4'!}:

\smallskip
\begin{verbatim}[commandchars=\\\{\}]
=>  2::3::\textbf{let} l = [] \textbf{in let} f x = x + 1 \textbf{in let} r = \underline{3 + 1} \textbf{in} r::map f l
=>  2::3::\textbf{let} l = [] \textbf{in let} f x = x + 1 \textbf{in let} r = 4 \textbf{in} \underline{r}::map f l
=>  2::3::\textbf{let} l = [] \textbf{in let} f x = x + 1 \textbf{in} 4::map \underline{f} l
=>  2::3::\textbf{let} l = [] \textbf{in} 4::\underline{map (\textbf{fun} x -> x + 1)} l
=>  2::3::\textbf{let} l = [] \textbf{in}
      4::\underline{(\textbf{let} f x = x + 1 \textbf{in function} [] -> [] | a::l -> \textbf{let} r = f a \textbf{in} r::map f l)} l
=>  2::3::\textbf{let} l = [] \textbf{in}
      4::\underline{(\textbf{function} [] -> [] | a::l -> \textbf{let} f x = x + 1 \textbf{in let} r = f a \textbf{in} r::map f l)} l
=>  2::3::\textbf{let} l = [] \textbf{in}
=>  2::3::4::
      \underline{(\textbf{function} [] -> [] | a::l -> \textbf{let} f x = x + 1 \textbf{in let} r = f a \textbf{in} r::map f l ) []}
\end{verbatim}
\smallskip

\noindent These searching mechanisms were arrived at through conjecture about and exploration of the most likely useful tools. It remains to be seen what the best interface for our interpreter or debugger will be. We discuss one promising option now -- a mechanism for annotating source code to be interpreted, leaving the rest of the program running natively at full speed.

\subsection{An interface for debugging}

In this section, we propose an unimplemented interface for debugging which uses the interpreter in a novel way, inserting it into a normal compiled OCaml program. In Section 3.6, we mentioned a system for expanding {\small\texttt{\textbf{external}}} definitions into shims for calling into C code:

\begin{verbatim}[commandchars=\\\{\}]
[%%auto \textbf{external} string_of_float : float -> string = "%string_of_float"]
\end{verbatim}

These {\small\verb![%%<name> <contents>]!} annotations are in fact a standard feature of recent versions of OCaml, called PPX annotations. They may be attached to any parse tree node. When OCaml compiles a file, it resolves such annotations by calling external processes (PPX extensions) which replace them with a parse tree. Once all annotations are resolved, the source is typechecked and compiled as usual. And so, we may use this mechanism to allow an OCaml program to be compiled as usual, except that the part we wish to debug (and so be interpreted) is marked with an {\small\texttt{[@interpret]}} annotation. This time, instead of communicating between OCaml and C code, we are communicating between the interpreted OCaml code and compiled OCaml code.

The use of {\small\texttt{[@interpret]}} annotations to control which parts of the code are executed natively and which parts are interpreted (and so have their steps of evaluation displayed on the screen) was motivated by our observation that a tool like \textsf{OCamli} on its own would not fulfil our usability needs, in particular  our requirement that we must get it ``inside'' the build process. The improvement in speed (by interpreting only what we need to debug) is a side-effect, but a pleasant one. Is another such side-effect of this mechanism a natural and pliable human interface for debugging? If it is, we may be within sight of achieving our original aim of a usable debugger. The interface would be:

\begin{enumerate}
  \item Notice that a misbehaviour is occurring.
  \item Knowing or speculating upon the location of the root cause, insert one or more appropriate {\small\texttt{[@interpret]}} annotations in the code.
  \item Recompile and run the code. The evaluation of the parts chosen will be shown.
  \item If the source or nature of the bug is now clear:
   \begin{enumerate}
     \item Change the source to fix the bug.
     \item Build and run again and inspect the output to be sure it is fixed.
     \item Remove the {\small\texttt{[@interpret]}} annotation(s).
   \end{enumerate}
  \item If the source or nature of the bug is not yet clear, due to a wrong or insufficient choice of {\small\texttt{[@interpret]}} annotations, return to step 2.
\end{enumerate}

\noindent Let us explore the design space of such debugging annotations. Consider the following possibilities, for instance:

\begin{description}

\item [{\small\texttt{\textmd{[@interpret]}}}] The piece of code annotated is interpreted, but functions it calls into are not. For example, consider the following buggy function on lists:

\begin{Verbatim}[commandchars=\\\{\}]
\textbf{let rec} pairs f a l =
  \textbf{match} l \textbf{with}
    [] -> rev a
  | [_] -> []
  | h::h'::t -> pairs f (f h h'::a) t

\textbf{let} x = pairs ( + ) [] [1; 2; 3; 4]
\end{Verbatim}

\noindent It is supposed to take, for example {\small\texttt{[1; 2; 3; 4]}} to {\small\texttt{[1 + 2; 2 + 3; 3 + 4]}} if the function {\small\texttt{f}} is addition. The input {\small\texttt{a}} is an accumulator to make it tail-recursive. There are two bugs in this version above. Firstly, for the case of the single-item list, the result should be {\small\texttt{rev a}}, as for the empty list. Secondly, the last case should read {\small\verb!pairs f (f h h'::a) (h'::t)!}.

We can add an {\small\texttt{[@interpret]}} annotation on to the outer invocation of {\small\texttt{pairs}}:

\begin{Verbatim}[commandchars=\\\{\}]
\textbf{let rec} pairs f a l =
  \textbf{match} l \textbf{with}
    [] -> rev a
  | [_] -> []
  | h::h'::t -> pairs f (f h h'::a) t

\textbf{let} x = [@interpret] pairs ( + ) [] [1; 2; 3; 4]
\end{Verbatim}

\noindent Now, upon compiling the program the interpreter is embedded, and all the code inside {\small\texttt{pairs}} is interpreted, but external calls (for example to {\small\texttt{rev}}) are native, and so elided from the output. We think this is the sensible default, both for elision of information and elision of computation. The output upon running the program would be the following:

\begin{Verbatim}[commandchars=\\\{\}]
   pairs ( + ) [] [1; 2; 3; 4]
\{matches h::h'::t\}
=> pairs ( + ) [3] [3; 4]
\{matches h::h'::t\}
=> pairs ( + ) [7; 3] []
\{matches []\}
=> [3; 7]
\end{Verbatim}

\noindent The first three lines are generated from the pairs function call itself, the last line from the returned value. Note that the default elision also does not show the details of the matches in as much detail as the output of \textsf{OCamli} we showed in Section 2.1. The bug is plain to see, so we correct it:

\begin{Verbatim}[commandchars=\\\{\}]
\textbf{let rec} pairs f a l =
  \textbf{match} l \textbf{with}
    [] -> rev a
  | [_] -> []
  | h::h'::t -> pairs f (f h h'::a) (h'::t)

\textbf{let} x = [@interpret] pairs ( + ) [] [1; 2; 3; 4]
\end{Verbatim}

We compile the code again, with the annotation in the same place, and try again:

\begin{Verbatim}[commandchars=\\\{\}]
   pairs ( + ) [] [1; 2; 3; 4]
\{matches h::h'::t\}
=> pairs ( + ) [3] [2; 3; 4]
\{matches h::h'::t\}
=> pairs ( + ) [5; 3] [3; 4]
\{matches h::h'::t\}
=> pairs ( + ) [7; 5; 3] [4]
\{matches [_]\}
=> []
\end{Verbatim}

\noindent Still there is a bug. Since the accumulator looks close to correct during evaluation, only to see the output disappear at the last moment, we deduce it must be the match case {\small\texttt{[\_]}} which is wrong, and we correct it:

\begin{Verbatim}[commandchars=\\\{\}]
\textbf{let rec} pairs f a l =
  \textbf{match} l \textbf{with}
    [] | [_] -> rev a
  | h::h'::t -> pairs f (f h h'::a) (h'::t)

\textbf{let} x = [@interpret] pairs ( + ) [] [1; 2; 3; 4]
\end{Verbatim}

Here is the final, correct output:

\begin{Verbatim}[commandchars=\\\{\}]
   pairs ( + ) [] [1; 2; 3; 4]
\{matches h::h'::t\}
=> pairs ( + ) [3] [2; 3; 4]
\{matches h::h'::t\}
=> pairs ( + ) [5; 3] [3; 4]
\{matches h::h'::t\}
=> pairs ( + ) [7; 5; 3] [4]
\{matches [_]\}
=> [3; 5; 7]
\end{Verbatim}

Now, we can remove our annotation:

\begin{Verbatim}[commandchars=\\\{\}]
\textbf{let rec} pairs f a l =
  \textbf{match} l \textbf{with}
    [] | [_] -> rev a
  | h::h'::t -> pairs f (f h h'::a) (h'::t)

\textbf{let} x = pairs ( + ) [] [1; 2; 3; 4]
\end{Verbatim}

Our debugging is complete.

\item [{\small\texttt{\textmd{[@interpret-all]}}}] The piece of code annotated is interpreted, and so is every function it calls, if available (the default action of {\small\texttt{[@interpret]}} is that calls into other modules are native). 

\item [{\small\texttt{\textmd{[@interpret-logto]}}}] The output is not written to standard output or standard error, but appended to a file. This can be used to separate the output of several annotations, or several runs of the same program, or as a crude logging mechanism.

\item [{\small\texttt{\textmd{[@interpret-env]}}}] Interpret only if an environment variable is set, otherwise run natively. This would allow code to remain unaltered after debugging, leaving the annotation in place in case the bug is not really fixed.

\item [{\small\texttt{\textmd{[@interpret-sub]}}}] Pause the program at the expression given, printing the current expression and allowing the programmer to substitute their own.

\item [{\small\texttt{\textmd{[@show-only]}}}] Simply show a given expression, but run it natively. This is a way of adding a generic printer to OCaml (like Java's {\small\texttt{toString}} or Haskell's {\small\texttt{show}}).

\item [{\small\texttt{\textmd{[@interpret-matching]}}}] Give a search term (like \textsf{OCamli}) and show only those lines of the evaluation when interpreting.

\item [{\small\texttt{\textmd{[@interpret-n <n>]}}}] Show only {\small\texttt{<n>}} times through this code point. After that, be silent whilst continuing to run.

\item [{\small\texttt{\textmd{[@exit-after <n>]}}}] Show only {\small\texttt{<n>}} times through this code point. After that, exit.
 
\item [{\small\texttt{\textmd{[@interpret-interactive <n>]}}}] Upon interpreting, dump into an \textsf{OCamli}-like interface which acts as an interactive debugger, setting and clearing breakpoints and so on.

\end{description}

\noindent We can see that many well-known mechanisms of debugging, such as breakpointing, find a new home here. The approach is in general a low-impact one: the programmer need use only the parts of the debugging system they wish to, or which suit their mental model or debugging style. We hope that this makes the debugger more likely to be used by more people.

It is possible to imagine other interface models, of course. We could, for example, produce a REPL-like program which is able to show steps of evaluation in addition to its usual functionality, for learning or testing, or light debugging. But the annotation-based one we have alighted upon seems promising, so we persevere with it for now.

It is important to point out that for this scheme to work, there would need to be a mechanism for types to be known at the time the annotation is expanded, such that values can be reliably used across the compiled/interpreted boundary. This might require a typed analog of the PPX system, where the annotation is expanded after typechecking has occured. But this is a relatively minor modification to the OCaml compiler.

\textbf{Summary}\hspace{3.5mm} We have demonstrated how to get our interpreter inside the build process of programs written in, or partly in, OCaml, and noted that this helps us achieve many of the aims of making a debugger which is available whenever the programmer needs it. We have seen how selective interpretation could lead to much more reasonable running times for programs we wish to debug by reducing the part which need be interpreted.

\section{Evaluation}

We have looked again at the problem of debugging, identified what we believe to be a debugger design which may lead to more people using debugging tools, and built a proof-of-concept of part of it for the functional language OCaml. By what criteria do we measure success? The primary measure, of course, is whether the tool, once finished, is widely used. Quantitively, we can measure two things: how many people use the tool in preference to another, and how many in preference to nothing at all or more often than they used their previous tool. Does it replace or merely complement other tools? 

Our plan was to look at the literature and practice of debugging, and try to discern those qualities which separate a debugger which is useful and used from one which lies unused. This was motivated by observing the widespread feeling amongst programmers that using a debugger would be more common if only it were more easily applicable to their problems. That is to say, the feeling that there is nothing fundamentally impossible about producing a widely used debugger. We looked at this in the context of functional programming especially, working on the assumption that functional programming is different enough from imperative programming that there are likely to be some differences in the debugging process.

We have identified what we believe to be the key requirements for a usable debugger -- that it should be available all the time, whenever the programmer needs it, and that it should be sufficiently flexible so as to be unobtrusive when not in use. There are, of course, many other requirements for a good debugger. But we have claimed that without the accessibility requirement being fulfilled, the rest is in vain. We chose the radical approach to these key requirements: to build a naive, step-by-step  interpreter. The supposed advantages were that this would result in accessibility-by-default, that there would be no information loss (since there is no compilation process), and that the obvious downside of interpretation -- slowness -- could be mitigated.

\textsf{OCamli} was written, firstly, to answer the question ``What would an interpreter for OCaml look like?'', and secondly to begin to explore the design space of visualizing (and thereby debugging) OCaml programs. However, it does not pass our tests of what a good debugger would look like -- and not only because it does not yet support the whole language. Let us suppose now that \textsf{OCamli} were to be finished to support the whole of OCaml. What would it still lack, with regard to our principles and, in particular, the tests we set out in Section \ref{tests}? \textsf{OCamli}, as presently constructed, does not meet the most important of the tests. It cannot be used as an alternative to the compiler except for the simplest of projects, and cannot be used with mixed C/OCaml code. These are design flaws, and impact upon usability, in particular the notion of accessibility we have been concerned with. 

\textsf{OCamli}, then, is an interesting exposition of our central idea of debugging-by-interpretation, but flawed with regard to usability, both in its choice of interface, and by dint of its failure to address fully the notion of accessibility. We have identified a solution to this problems, a mechanism by which the interpreter may be placed in an executable, alongside native code. We believe this could solve the remaining problems.

\section{Conclusions and future work}
We have looked again at the history and present practice of debugging and tried to identify the essential characteristics which separate debuggers which are likely to be used from those which are likely not to be. Working from these principles, we have described a design for, and early prototype of, a new debugger for the functional language OCaml based on the concept of direct interpretation, and a design for a mechanism for it to be embedded into the build process in such a way that it is always available. We believe it to be promising, but it is too early to say if it really represents a significant step forward -- the problem of debugging is old and intransigent.

As we have discussed, the ultimate test is, of course, whether anyone uses it. So, simply put, our most important item of future work is to provide a complete implementation as a concrete way of supporting (or undermining) our thesis.

Some of the technical mechanisms we have used to create this debugger are rather specific to OCaml -- to what extent might the insights gained be useful in other languages? It would be interesting to see if this mixture of interpretation and compilation can be applied elsewhere. We have, thus far, made no attempt at preserving the time or space complexity of programs under interpretation, even when the intermediate steps are not shown. Is it possible to interpret programs in such a way that we can give a guarantee about the complexity?

In addition, we have not yet formalized certain parts of the implementation, for example how to match the OCaml compiler's actual order of evaluation, in cases where it is not specified in the language -- an important part of being able to reproduce bugs using the debugger. There will doubtless be several difficult little issues like this.

The kind of diagrams our interpreter draws are also useful for teaching, that is to say testing out little programs a student is writing rather than debugging large codebases. It would be interesting to look at how exactly learning to program and debugging are intertwined or equivalent tasks, and see if our interpreter helps beginning students. Besides teaching and debugging, having a small-step interpreter readily available for a language, especially one which ranks equally with the compiler and can be mixed with it at will, may have more uses which we have yet to discover.

\nocite{*}
\bibliographystyle{eptcs}
\bibliography{ocamli}

\begin{thebibliography}{10}
\providecommand{\bibitemdeclare}[2]{}
\providecommand{\surnamestart}{}
\providecommand{\surnameend}{}
\providecommand{\urlprefix}{Available at }
\providecommand{\url}[1]{\texttt{#1}}
\providecommand{\href}[2]{\texttt{#2}}
\providecommand{\urlalt}[2]{\href{#1}{#2}}
\providecommand{\doi}[1]{doi:\urlalt{http://dx.doi.org/#1}{#1}}
\providecommand{\bibinfo}[2]{#2}

\bibitemdeclare{inproceedings}{visual-miranda}
\bibitem{visual-miranda}
\bibinfo{author}{Mikhail \surnamestart Auguston\surnameend} \&
  \bibinfo{author}{Juris \surnamestart Reinfields\surnameend}
  (\bibinfo{year}{1994}): \emph{\bibinfo{title}{A {Visual} {Miranda}
  {Machine}}}.
\newblock In: {\sl \bibinfo{booktitle}{Software Education Conference, 1994.
  Proceedings.}}, \bibinfo{organization}{IEEE}, pp. \bibinfo{pages}{198--203},
  \doi{10.1109/SEDC.1994.475337}.

\bibitemdeclare{inproceedings}{balzer1969exdams}
\bibitem{balzer1969exdams}
\bibinfo{author}{R.M. \surnamestart Balzer\surnameend} (\bibinfo{year}{1969}):
  \emph{\bibinfo{title}{EXDAMS -- Extendable Debugging And Monitoring System}}.
\newblock In: {\sl \bibinfo{booktitle}{Proceedings of the May 14-16, 1969,
  {S}pring {J}oint {C}omputer {C}onference}}, \bibinfo{series}{Memorandum (Rand
  Corporation)}, \bibinfo{organization}{ACM}, \bibinfo{publisher}{Rand
  Corporation}, pp. \bibinfo{pages}{567--580}, \doi{10.1145/1476793.1476881}.

\bibitemdeclare{article}{brady1968writing}
\bibitem{brady1968writing}
\bibinfo{author}{Paul~T \surnamestart Brady\surnameend} (\bibinfo{year}{1968}):
  \emph{\bibinfo{title}{Writing an online debugging program for the experienced
  user}}.
\newblock {\sl \bibinfo{journal}{Communications of the ACM}}
  \bibinfo{volume}{11}(\bibinfo{number}{6}), pp. \bibinfo{pages}{423--427},
  \doi{10.1145/363347.363388}.

\bibitemdeclare{inproceedings}{burstall1980hope}
\bibitem{burstall1980hope}
\bibinfo{author}{Rod~M \surnamestart Burstall\surnameend},
  \bibinfo{author}{David~B \surnamestart MacQueen\surnameend} \&
  \bibinfo{author}{Donald~T \surnamestart Sannella\surnameend}
  (\bibinfo{year}{1980}): \emph{\bibinfo{title}{HOPE: An experimental
  applicative language}}.
\newblock In: {\sl \bibinfo{booktitle}{Proceedings of the 1980 ACM conference
  on LISP and functional programming}}, \bibinfo{organization}{ACM}, pp.
  \bibinfo{pages}{136--143}, \doi{10.1145/800087.802799}.

\bibitemdeclare{inproceedings}{chargueraud2018jsexplain}
\bibitem{chargueraud2018jsexplain}
\bibinfo{author}{Arthur \surnamestart Chargu{\'e}raud\surnameend},
  \bibinfo{author}{Alan \surnamestart Schmitt\surnameend} \&
  \bibinfo{author}{Thomas \surnamestart Wood\surnameend}
  (\bibinfo{year}{2018}): \emph{\bibinfo{title}{JSExplain: A Double Debugger
  for JavaScript}}.
\newblock In: {\sl \bibinfo{booktitle}{The Web Conference 2018}}, pp.
  \bibinfo{pages}{1--9}, \doi{10.1145/3184558.3185969}.

\bibitemdeclare{inproceedings}{chitil2002transforming}
\bibitem{chitil2002transforming}
\bibinfo{author}{Olaf \surnamestart Chitil\surnameend}, \bibinfo{author}{Colin
  \surnamestart Runciman\surnameend} \& \bibinfo{author}{Malcolm \surnamestart
  Wallace\surnameend} (\bibinfo{year}{2002}):
  \emph{\bibinfo{title}{Transforming Haskell for tracing}}.
\newblock In: {\sl \bibinfo{booktitle}{Symposium on Implementation and
  Application of Functional Languages}}, \bibinfo{organization}{Springer}, pp.
  \bibinfo{pages}{165--181}, \doi{10.1007/3-540-44854-3_11}.

\bibitemdeclare{inproceedings}{Clements2001}
\bibitem{Clements2001}
\bibinfo{author}{John \surnamestart Clements\surnameend},
  \bibinfo{author}{Matthew \surnamestart Flatt\surnameend} \&
  \bibinfo{author}{Matthias \surnamestart Felleisen\surnameend}
  (\bibinfo{year}{2001}): \emph{\bibinfo{title}{Modeling an Algebraic
  Stepper}}.
\newblock In: {\sl \bibinfo{booktitle}{Proceedings of the 10th European
  Symposium on Programming Languages and Systems}}, \bibinfo{series}{ESOP '01},
  \bibinfo{publisher}{Springer-Verlag}, \bibinfo{address}{London, UK, UK}, pp.
  \bibinfo{pages}{320--334}, \doi{10.1007/3-540-45309-1_21}.

\bibitemdeclare{article}{cong2016implementing}
\bibitem{cong2016implementing}
\bibinfo{author}{Youyou \surnamestart Cong\surnameend} \&
  \bibinfo{author}{Kenichi \surnamestart Asai\surnameend}
  (\bibinfo{year}{2016}): \emph{\bibinfo{title}{Implementing a stepper using
  delimited continuations}}.
\newblock {\sl \bibinfo{journal}{Software Science}} \bibinfo{volume}{39}, pp.
  \bibinfo{pages}{42--54}, \doi{10.29007/l2wb}.

\bibitemdeclare{}{fpcomplete}
\bibitem{fpcomplete}
\bibinfo{author}{Aaron \surnamestart Contorer\surnameend}
  (\bibinfo{year}{2015}): \emph{\bibinfo{title}{What do Haskellers want? Over a
  thousand users tell us.}}
\newblock
  \urlprefix\url{https://www.fpcomplete.com/blog/2015/05/thousand-user-haskell-survey}.

\bibitemdeclare{inproceedings}{crew1997astlog}
\bibitem{crew1997astlog}
\bibinfo{author}{Roger~F \surnamestart Crew\surnameend} et~al.
  (\bibinfo{year}{1997}): \emph{\bibinfo{title}{ASTLOG: A Language for
  Examining Abstract Syntax Trees.}}
\newblock In: {\sl \bibinfo{booktitle}{DSL}}, \bibinfo{volume}{97}, pp.
  \bibinfo{pages}{18--18}.
\newblock
  \urlprefix\url{https://www.usenix.org/legacy/publications/library/proceedings/dsl97/full_papers/crew/crew.pdf}.

\bibitemdeclare{article}{devanbu1999genoa}
\bibitem{devanbu1999genoa}
\bibinfo{author}{Premkumar~T \surnamestart Devanbu\surnameend}
  (\bibinfo{year}{1999}): \emph{\bibinfo{title}{GENOA -- a customizable,
  front-end-retargetable source code analysis framework}}.
\newblock {\sl \bibinfo{journal}{ACM Transactions on Software Engineering and
  Methodology (TOSEM)}} \bibinfo{volume}{8}(\bibinfo{number}{2}), pp.
  \bibinfo{pages}{177--212}, \doi{10.1145/304399.304402}.

\bibitemdeclare{article}{eisenstadt1997my}
\bibitem{eisenstadt1997my}
\bibinfo{author}{Marc \surnamestart Eisenstadt\surnameend}
  (\bibinfo{year}{1997}): \emph{\bibinfo{title}{My hairiest bug war stories}}.
\newblock {\sl \bibinfo{journal}{Communications of the ACM}}
  \bibinfo{volume}{40}(\bibinfo{number}{4}), pp. \bibinfo{pages}{30--37},
  \doi{10.1145/248448.248456}.

\bibitemdeclare{inproceedings}{evans1966line}
\bibitem{evans1966line}
\bibinfo{author}{Thomas~G \surnamestart Evans\surnameend} \&
  \bibinfo{author}{D~Lucille \surnamestart Darley\surnameend}
  (\bibinfo{year}{1966}): \emph{\bibinfo{title}{On-line debugging techniques: a
  survey}}.
\newblock In: {\sl \bibinfo{booktitle}{Proceedings of the November 7-10, 1966,
  {F}all {J}oint {C}omputer {C}onference}}, \bibinfo{organization}{ACM}, pp.
  \bibinfo{pages}{37--50}, \doi{10.1145/1464291.1464295}.

\bibitemdeclare{inproceedings}{racket}
\bibitem{racket}
\bibinfo{author}{Matthias \surnamestart Felleisen\surnameend},
  \bibinfo{author}{Robert~Bruce \surnamestart Findler\surnameend},
  \bibinfo{author}{Matthew \surnamestart Flatt\surnameend},
  \bibinfo{author}{Shriram \surnamestart Krishnamurthi\surnameend},
  \bibinfo{author}{Eli \surnamestart Barzilay\surnameend}, \bibinfo{author}{Jay
  \surnamestart McCarthy\surnameend} \& \bibinfo{author}{Sam \surnamestart
  Tobin-Hochstadt\surnameend} (\bibinfo{year}{2015}): \emph{\bibinfo{title}{The
  racket manifesto}}.
\newblock In: {\sl \bibinfo{booktitle}{LIPIcs-Leibniz International Proceedings
  in Informatics}}, \bibinfo{volume}{32}, \bibinfo{organization}{Schloss
  Dagstuhl-Leibniz-Zentrum fuer Informatik},
  \doi{10.4230/LIPIcs.SNAPL.2015.113}.

\bibitemdeclare{article}{drscheme}
\bibitem{drscheme}
\bibinfo{author}{Robert~Bruce \surnamestart Finder\surnameend},
  \bibinfo{author}{John \surnamestart Clements\surnameend},
  \bibinfo{author}{Cormac \surnamestart Flanagan\surnameend},
  \bibinfo{author}{Matthew \surnamestart Flatt\surnameend},
  \bibinfo{author}{Shriram \surnamestart Krishnamurthi\surnameend},
  \bibinfo{author}{Paul \surnamestart Steckler\surnameend} \&
  \bibinfo{author}{Matthias \surnamestart Felleisen\surnameend}
  (\bibinfo{year}{2002}): \emph{\bibinfo{title}{{DrScheme}: {A} {Programming}
  {Environment} for {Scheme}}}.
\newblock {\sl \bibinfo{journal}{Journal of Functional Programming}}
  \bibinfo{volume}{12}(\bibinfo{number}{2}), pp. \bibinfo{pages}{159--182},
  \doi{10.1017/S0956796801004208}.

\bibitemdeclare{}{fsharpdebugging}
\bibitem{fsharpdebugging}
\emph{\bibinfo{title}{Visual Studio Docs: Debugging F\#}}.
\newblock
  \urlprefix\url{http://docs.microsoft.com/en-us/visualstudio/debugger/debugging-f-hash}.

\bibitemdeclare{article}{Gill00debugginghaskell}
\bibitem{Gill00debugginghaskell}
\bibinfo{author}{Andy \surnamestart Gill\surnameend} (\bibinfo{year}{2000}):
  \emph{\bibinfo{title}{{Debugging} {Haskell} by {Observing} {Intermediate}
  {Data} {Structures}}}.
\newblock {\sl \bibinfo{journal}{Electr. Notes Theor. Comput. Sci.}}
  \bibinfo{volume}{41}(\bibinfo{number}{1}), p.~\bibinfo{pages}{1},
  \doi{10.1016/S1571-0661(05)80538-9}.

\bibitemdeclare{article}{miracalc}
\bibitem{miracalc}
\bibinfo{author}{Doug \surnamestart Goldson\surnameend} (\bibinfo{year}{1993}):
  \emph{\bibinfo{title}{A {Symbolic} {Calculator} for {Non-Strict} {Functional}
  {Programs}}}.
\newblock {\sl \bibinfo{journal}{The Computer Journal}}
  \bibinfo{volume}{37}(\bibinfo{number}{3}), pp. \bibinfo{pages}{177--187},
  \doi{10.1093/comjnl/37.3.177}.

\bibitemdeclare{inproceedings}{grishman1970debugging}
\bibitem{grishman1970debugging}
\bibinfo{author}{Ralph \surnamestart Grishman\surnameend}
  (\bibinfo{year}{1970}): \emph{\bibinfo{title}{The debugging system AIDS}}.
\newblock In: {\sl \bibinfo{booktitle}{Proceedings of the May 5-7, 1970,
  {S}pring {J}oint {C}omputer {C}onference}}, \bibinfo{organization}{ACM}, pp.
  \bibinfo{pages}{59--64}, \doi{10.1145/1476936.1476952}.

\bibitemdeclare{inproceedings}{griswold1996fast}
\bibitem{griswold1996fast}
\bibinfo{author}{William~G \surnamestart Griswold\surnameend},
  \bibinfo{author}{Darren~C \surnamestart Atkinson\surnameend} \&
  \bibinfo{author}{Collin \surnamestart McCurdy\surnameend}
  (\bibinfo{year}{1996}): \emph{\bibinfo{title}{Fast, flexible syntactic
  pattern matching and processing}}.
\newblock In: {\sl \bibinfo{booktitle}{Program Comprehension, 1996,
  Proceedings., Fourth Workshop on}}, \bibinfo{organization}{IEEE}, pp.
  \bibinfo{pages}{144--153}, \doi{10.1109/WPC.1996.501129}.

\bibitemdeclare{article}{hall}
\bibitem{hall}
\bibinfo{author}{Cordelia~V \surnamestart Hall\surnameend} \&
  \bibinfo{author}{John~T \surnamestart O'Donnell\surnameend}
  (\bibinfo{year}{1985}): \emph{\bibinfo{title}{Debugging in a side effect free
  programming environment}}.
\newblock {\sl \bibinfo{journal}{ACM SIGPLAN Notices}}
  \bibinfo{volume}{20}(\bibinfo{number}{7}), pp. \bibinfo{pages}{60--68},
  \doi{10.1145/17919.806827}.

\bibitemdeclare{article}{halpern1965computer}
\bibitem{halpern1965computer}
\bibinfo{author}{Mark \surnamestart Halpern\surnameend} (\bibinfo{year}{1965}):
  \emph{\bibinfo{title}{Computer programming: the debugging epoch opens}}.
\newblock {\sl \bibinfo{journal}{Computers and Automation}}
  \bibinfo{volume}{14}(\bibinfo{number}{11}), pp. \bibinfo{pages}{28--31}.

\bibitemdeclare{article}{halpern2005assertive}
\bibitem{halpern2005assertive}
\bibinfo{author}{Mark \surnamestart Halpern\surnameend} (\bibinfo{year}{2005}):
  \emph{\bibinfo{title}{Assertive Debugging: Correcting Software As If We Meant
  It -- Assertive debugging is a new way to make embedded systems ensure their
  own health by having your code monitor itself.}}
\newblock {\sl \bibinfo{journal}{Embedded Systems Programming}}
  \bibinfo{volume}{18}(\bibinfo{number}{6}), pp. \bibinfo{pages}{28--35}.

\bibitemdeclare{article}{hamlet1983}
\bibitem{hamlet1983}
\bibinfo{author}{Dick \surnamestart Hamlet\surnameend} (\bibinfo{year}{1983}):
  \emph{\bibinfo{title}{Debugging "Level": Step-wise Debugging}}.
\newblock {\sl \bibinfo{journal}{SIGPLAN Not.}}
  \bibinfo{volume}{18}(\bibinfo{number}{8}), pp. \bibinfo{pages}{4--8},
  \doi{10.1145/1006142.1006150}.

\bibitemdeclare{}{ocaml}
\bibitem{ocaml}
\bibinfo{author}{Xavier \surnamestart Leroy\surnameend},
  \bibinfo{author}{Damien \surnamestart Doligez\surnameend},
  \bibinfo{author}{Alain \surnamestart Frisch\surnameend},
  \bibinfo{author}{Jacques \surnamestart Garrigue\surnameend},
  \bibinfo{author}{Didier \surnamestart R{\'e}my\surnameend} \&
  \bibinfo{author}{J{\'e}r{\^o}me \surnamestart Vouillon\surnameend}
  (\bibinfo{year}{2018}): \emph{\bibinfo{title}{The {OCaml} {Language}}}.
\newblock \urlprefix\url{http://ocaml.org/}.

\bibitemdeclare{inproceedings}{marlow}
\bibitem{marlow}
\bibinfo{author}{Simon \surnamestart Marlow\surnameend},
  \bibinfo{author}{Jos{\'e} \surnamestart Iborra\surnameend},
  \bibinfo{author}{Bernard \surnamestart Pope\surnameend} \&
  \bibinfo{author}{Andy \surnamestart Gill\surnameend} (\bibinfo{year}{2007}):
  \emph{\bibinfo{title}{A {Lightweight} {Interactive} {Debugger} for
  {Haskell}}}.
\newblock In: {\sl \bibinfo{booktitle}{Proceedings of the ACM SIGPLAN Workshop
  on Haskell Workshop}}, \bibinfo{series}{Haskell '07},
  \bibinfo{publisher}{ACM}, \bibinfo{address}{New York, NY, USA}, pp.
  \bibinfo{pages}{13--24}, \doi{10.1145/1291201.1291204}.

\bibitemdeclare{}{clr}
\bibitem{clr}
\bibinfo{author}{Erik \surnamestart Meijer\surnameend} \& \bibinfo{author}{John
  \surnamestart Gough\surnameend} (\bibinfo{year}{2012}):
  \emph{\bibinfo{title}{Technical overview of the Common Language Runtime,
  2000}}.
\newblock \urlprefix\url{https://research. microsoft.
  com/en-us/um/people/emeijer/papers/clr. pdf}.

\bibitemdeclare{article}{milner}
\bibitem{milner}
\bibinfo{author}{Robin \surnamestart Milner\surnameend} (\bibinfo{year}{1983}):
  \emph{\bibinfo{title}{How ML evolved}}.
\newblock {\sl \bibinfo{journal}{Polymorphism: The ML/LCF/Hope Newsletter}}
  \bibinfo{volume}{1}.

\bibitemdeclare{}{ocamli}
\bibitem{ocamli}
\emph{\bibinfo{title}{The {OCamli} {Interpreter}}}.
\newblock \urlprefix\url{http://github.com/johnwhitington/ocamli}.

\bibitemdeclare{article}{winhipe}
\bibitem{winhipe}
\bibinfo{author}{Crist{\'o}bal \surnamestart Pareja-Flores\surnameend},
  \bibinfo{author}{Jaime \surnamestart Urquiza-Fuentes\surnameend} \&
  \bibinfo{author}{J.~{\'A}ngel \surnamestart
  Vel{\'a}zquez-Iturbide\surnameend} (\bibinfo{year}{2007}):
  \emph{\bibinfo{title}{{WinHIPE}: {An} {IDE} for {Functional} {Programming}
  {Based} on {Rewriting} and {Visualization}}}.
\newblock {\sl \bibinfo{journal}{{ACM} {SIGPLAN} {Notices}}}
  \bibinfo{volume}{42}(\bibinfo{number}{3}), pp. \bibinfo{pages}{14--23},
  \doi{10.1145/1273039.1273042}.

\bibitemdeclare{inproceedings}{Parnin:2011:ADT:2001420.2001445}
\bibitem{Parnin:2011:ADT:2001420.2001445}
\bibinfo{author}{Chris \surnamestart Parnin\surnameend} \&
  \bibinfo{author}{Alessandro \surnamestart Orso\surnameend}
  (\bibinfo{year}{2011}): \emph{\bibinfo{title}{Are Automated Debugging
  Techniques Actually Helping Programmers?}}
\newblock In: {\sl \bibinfo{booktitle}{Proceedings of the 2011 International
  Symposium on Software Testing and Analysis}}, \bibinfo{series}{ISSTA '11},
  \bibinfo{publisher}{ACM}, \bibinfo{address}{New York, NY, USA}, pp.
  \bibinfo{pages}{199--209}, \doi{10.1145/2001420.2001445}.

\bibitemdeclare{article}{paul1994framework}
\bibitem{paul1994framework}
\bibinfo{author}{Santanu \surnamestart Paul\surnameend} \&
  \bibinfo{author}{Atul \surnamestart Prakash\surnameend}
  (\bibinfo{year}{1994}): \emph{\bibinfo{title}{A framework for source code
  search using program patterns}}.
\newblock {\sl \bibinfo{journal}{IEEE Transactions on Software Engineering}}
  \bibinfo{volume}{20}(\bibinfo{number}{6}), pp. \bibinfo{pages}{463--475},
  \doi{10.1109/32.295894}.

\bibitemdeclare{article}{PetredeQuincey}
\bibitem{PetredeQuincey}
\bibinfo{author}{Marian \surnamestart Petre\surnameend} \&
  \bibinfo{author}{Ed~\surnamestart de~Quincey\surnameend}
  (\bibinfo{year}{2006}): \emph{\bibinfo{title}{A {Gentle} {Overview} of
  {Software} {Visualization}}}.
\newblock {\sl \bibinfo{journal}{PPIG Newsletter}}, pp. \bibinfo{pages}{1--10}.

\bibitemdeclare{inproceedings}{Reeves95thecalculator}
\bibitem{Reeves95thecalculator}
\bibinfo{author}{Steve \surnamestart Reeves\surnameend}, \bibinfo{author}{Doug
  \surnamestart Goldson\surnameend}, \bibinfo{author}{Pat \surnamestart
  Fung\surnameend}, \bibinfo{author}{Mike \surnamestart Hopkins\surnameend} \&
  \bibinfo{author}{Richard \surnamestart Bornat\surnameend}
  (\bibinfo{year}{1994}): \emph{\bibinfo{title}{The {Calculator} {Project} --
  {Formal} {Reasoning} about {Programs}}}.
\newblock In: {\sl \bibinfo{booktitle}{Software Education Conference, 1994.
  Proceedings.}}, \bibinfo{organization}{IEEE}, pp. \bibinfo{pages}{166--173},
  \doi{10.1109/SEDC.1994.475332}.

\bibitemdeclare{inproceedings}{regelson1994debugging}
\bibitem{regelson1994debugging}
\bibinfo{author}{Elaine \surnamestart Regelson\surnameend} \&
  \bibinfo{author}{Andrew \surnamestart Anderson\surnameend}
  (\bibinfo{year}{1994}): \emph{\bibinfo{title}{Debugging Practices for Complex
  Legacy Software Systems.}}
\newblock In: {\sl \bibinfo{booktitle}{ICSM}}, pp. \bibinfo{pages}{137--143},
  \doi{10.1109/ICSM.1994.336781}.

\bibitemdeclare{}{libmonda}
\bibitem{libmonda}
\bibinfo{author}{Mark \surnamestart Shinwell\surnameend}:
  \emph{\bibinfo{title}{libmonda: Make OCaml native debugging awesome}}.
\newblock \urlprefix\url{http://mshinwell.github.io/libmonda/}.

\bibitemdeclare{article}{Sorva}
\bibitem{Sorva}
\bibinfo{author}{Juha \surnamestart Sorva\surnameend}, \bibinfo{author}{Ville
  \surnamestart Karavirta\surnameend} \& \bibinfo{author}{Lauri \surnamestart
  Malmi\surnameend} (\bibinfo{year}{2013}): \emph{\bibinfo{title}{A Review of
  Generic Program Visualization Systems for Introductory Programming
  Education}}.
\newblock {\sl \bibinfo{journal}{Transactions in Computing Education}}
  \bibinfo{volume}{13}(\bibinfo{number}{4}), pp. \bibinfo{pages}{15:1--15:64},
  \doi{10.1145/2490822}.

\bibitemdeclare{book}{stallman1981emacs}
\bibitem{stallman1981emacs}
\bibinfo{author}{Richard~M \surnamestart Stallman\surnameend}
  (\bibinfo{year}{1981}): \emph{\bibinfo{title}{EMACS the extensible,
  customizable self-documenting display editor}}.
\newblock \bibinfo{publisher}{ACM Books}, \doi{10.1145/800209.806466}.

\bibitemdeclare{article}{commonlisp}
\bibitem{commonlisp}
\bibinfo{author}{Guy \surnamestart Steele\surnameend} (\bibinfo{year}{1984}):
  \emph{\bibinfo{title}{Common LISP: the language}}.
\newblock {\sl \bibinfo{journal}{Digital Press}} \bibinfo{volume}{20}, p.
  \bibinfo{pages}{124}.

\bibitemdeclare{}{fsharp}
\bibitem{fsharp}
\bibinfo{author}{Don \surnamestart Syme\surnameend}: \emph{\bibinfo{title}{F
  Sharp at Microsoft Research}}.
\newblock
  \urlprefix\url{https://www.microsoft.com/en-us/research/project/f-at-microsoft-research/}.

\bibitemdeclare{phdthesis}{taylor-thesis}
\bibitem{taylor-thesis}
\bibinfo{author}{Jonathan~Paul \surnamestart Taylor\surnameend}
  (\bibinfo{year}{1996}): \emph{\bibinfo{title}{{Presenting} the {Lazy}
  {Evaluation} of {Functions}}}.
\newblock Ph.D. thesis, \bibinfo{school}{Queen Mary, University of London}.

\bibitemdeclare{article}{tolmach}
\bibitem{tolmach}
\bibinfo{author}{Andrew \surnamestart Tolmach\surnameend} \&
  \bibinfo{author}{Andrew~W. \surnamestart Appel\surnameend}
  (\bibinfo{year}{1995}): \emph{\bibinfo{title}{A {Debugger} for {Standard}
  {ML}}}.
\newblock {\sl \bibinfo{journal}{Journal of Functional Programming}}
  \bibinfo{volume}{5}, pp. \bibinfo{pages}{155--200},
  \doi{10.1017/S0956796800001313}.

\bibitemdeclare{article}{touretzky}
\bibitem{touretzky}
\bibinfo{author}{David~S. \surnamestart Touretzky\surnameend}
  (\bibinfo{year}{1989}): \emph{\bibinfo{title}{Visualizing {Evaluation} in
  {Applicative} {Languages}}}.
\newblock {\sl \bibinfo{journal}{Commun. ACM}}
  \bibinfo{volume}{35}(\bibinfo{number}{10}), pp. \bibinfo{pages}{49--59},
  \doi{10.1145/135239.135241}.

\bibitemdeclare{article}{zstep95}
\bibitem{zstep95}
\bibinfo{author}{David \surnamestart Ungar\surnameend}, \bibinfo{author}{Henry
  \surnamestart Lieberman\surnameend} \& \bibinfo{author}{Christopher
  \surnamestart Fry\surnameend} (\bibinfo{year}{1997}):
  \emph{\bibinfo{title}{{Debugging} and the {Experience} of {Immediacy}}}.
\newblock {\sl \bibinfo{journal}{Communications of the ACM}}
  \bibinfo{volume}{40}(\bibinfo{number}{4}), pp. \bibinfo{pages}{38--43},
  \doi{10.1145/248448.248457}.

\bibitemdeclare{inproceedings}{UrquizaFuentes}
\bibitem{UrquizaFuentes}
\bibinfo{author}{J.~\surnamestart Urquiza-Fuentes\surnameend} \&
  \bibinfo{author}{J.~A. \surnamestart Vel{\'a}zquez-Iturbide\surnameend}
  (\bibinfo{year}{2004}): \emph{\bibinfo{title}{A {Survey} of {Program}
  {Visualizations} for the {Functional} {Paradigm}}}.
\newblock In: {\sl \bibinfo{booktitle}{Proc. 3rd Program Visualization
  Workshop}}, pp. \bibinfo{pages}{2--9}.
\newblock \urlprefix\url{https://www.dcs.warwick.ac.uk/pvw04/p01.pdf}.

\bibitemdeclare{article}{survey2009}
\bibitem{survey2009}
\bibinfo{author}{Jaime \surnamestart Urquiza-Fuentes\surnameend} \&
  \bibinfo{author}{J.~{\'A}ngel \surnamestart
  Vel{\'a}zquez-Iturbide\surnameend} (\bibinfo{year}{2009}):
  \emph{\bibinfo{title}{A {Survey} of {Successful} {Evaluations} of {Program}
  {Visualisation} and {Algorithm} {Animation} {Systems}}}.
\newblock {\sl \bibinfo{journal}{ACM {T}ransactions on {C}omputing {E}ducation
  (TOCE)}} \bibinfo{volume}{9}(\bibinfo{number}{2}), p.~\bibinfo{pages}{9},
  \doi{10.1145/1538234.1538236}.

\bibitemdeclare{article}{Wadler}
\bibitem{Wadler}
\bibinfo{author}{Philip \surnamestart Wadler\surnameend}
  (\bibinfo{year}{1998}): \emph{\bibinfo{title}{Why {No} {One} {Uses}
  {Functional} {Languages}}}.
\newblock {\sl \bibinfo{journal}{SIGPLAN Not.}}
  \bibinfo{volume}{33}(\bibinfo{number}{8}), pp. \bibinfo{pages}{23--27},
  \doi{10.1145/286385.286387}.

\bibitemdeclare{inproceedings}{whitington}
\bibitem{whitington}
\bibinfo{author}{John \surnamestart Whitington\surnameend} \&
  \bibinfo{author}{Tom \surnamestart Ridge\surnameend} (\bibinfo{year}{2017}):
  \emph{\bibinfo{title}{Visualizing the Evaluation of Functional Programs for
  Debugging}}.
\newblock In: {\sl \bibinfo{booktitle}{6th Symposium on Languages, Applications
  and Technologies}}, {\sl \bibinfo{series}{OASIcs}}~\bibinfo{volume}{56},
  \bibinfo{publisher}{Schloss Dagstuhl--Leibniz-Zentrum fuer Informatik},
  \bibinfo{address}{Dagstuhl, Germany}, pp. \bibinfo{pages}{7:1--7:9},
  \doi{10.4230/OASIcs.SLATE.2017.7}.

\end{thebibliography}
\end{document}